\documentclass[sigconf,nonacm]{acmart}
\newcommand{\model}{\texttt{BiFair}}

\usepackage{algorithm}
\usepackage{algorithmic}
\usepackage{multicol}
\usepackage{multirow}
\usepackage{graphicx}
\usepackage{enumitem}

\AtBeginDocument{%
  }

\setcopyright{acmlicensed}
\copyrightyear{2018}
\acmYear{2018}
\acmDOI{XXXXXXX.XXXXXXX}
\acmConference[Conference acronym 'XX]{Make sure to enter the correct
  conference title from your rights confirmation email}{June 03--05,
  2018}{Woodstock, NY}
\acmISBN{978-1-4503-XXXX-X/2018/06}

\begin{document}

\title{BiFair: A Fairness-aware Training Framework for LLM-enhanced Recommender Systems via Bi-level Optimization}

\author{Jiaming Zhang}
\email{jm.zhang@zju.edu.cn}
\affiliation{%
  \institution{Zhejiang University}
  \city{Hangzhou}
  \country{China}
}
\author{Yuyuan Li}
\email{y2li@hdu.edu.cn}
\affiliation{%
  \institution{Hangzhou Dianzi University}
  \city{Hangzhou}
  \country{China}
}
\author{Yiqun Xu}
\email{abyss815643904@gmail.com}
\affiliation{%
  \institution{University of Electronic Science and Technology of China}
  \city{Chengdu}
  \country{China}
}
\author{Li Zhang}
\email{zhanglizl80@gmail.com}
\affiliation{%
  \institution{Zhejiang University}
  \city{Hangzhou}
  \country{China}
}
\author{Xiaohua Feng}
\email{fengxiaohua@zju.edu.cn}
\affiliation{%
  \institution{Zhejiang University}
  \city{Hangzhou}
  \country{China}
}
\author{Zhifei Ren}
\email{213231895@seu.edu.cn}
\affiliation{%
  \institution{Southeast University}
  \city{Nanjing}
  \country{China}
}
\author{Chaochao Chen}
\authornote{Corresponding author.}
\email{zjuccc@zju.edu.cn}
\affiliation{%
  \institution{Zhejiang University}
  \city{Hangzhou}
  \country{China}
}

\begin{abstract}
    Large Language Model-enhanced Recommender Systems (LLM-enhanced RSs) have emerged as a powerful approach to improving recommendation quality by leveraging LLMs to generate item representations.
    Despite these advancements, the integration of LLMs raises severe fairness concerns. Existing studies reveal that LLM-based RSs exhibit greater unfairness than traditional RSs, yet fairness issues in LLM-enhanced RSs remain largely unexplored.
    In this paper, our empirical study reveals that while LLM-enhanced RSs improve fairness across item groups, a significant fairness gap persists.
    Further enhancement remains challenging due to the architectural differences and varying sources of unfairness inherent in LLM-enhanced RSs.
    To bridge this gap, we first decompose unfairness into i) \textit{prior unfairness} in LLM-generated representations and ii) \textit{training unfairness} in recommendation models. 
    Then, we propose BiFair, a bi-level optimization-based fairness-aware training framework designed to mitigate both prior and training unfairness simultaneously. 
    BiFair optimizes two sets of learnable parameters: LLM-generated representations and a trainable projector in the recommendation model, using a two-level nested optimization process. 
    Additionally, we introduce an adaptive inter-group balancing mechanism, leveraging multi-objective optimization principles to dynamically balance fairness across item groups.
    Extensive experiments on three real-world datasets demonstrate that BiFair significantly mitigates unfairness and outperforms the state-of-the-art methods.
\end{abstract}




\maketitle

\section{INTRODUCTION}
Nowadays, Recommender Systems (RSs) have become integral to personalized information services.
Leveraging advanced algorithms and large-scale data, RSs help users efficiently identify content that meets their needs across various domains~\cite{turcotte2015news,bao2016intelligent,paparrizos2011machine}.
However, traditional RSs are often limited by their reliance on interaction data and static user behaviors, lacking the integration of comprehensive world knowledge and broader user behavior patterns.
Consequently, traditional RSs struggle to capture complex user preferences and deliver truly personalized experiences.

Recently, Large Language Models (LLMs)~\cite{achiam2023gpt,touvron2023llama,yang2024qwen2} have been widely adopted to enhance recommendation performance due to their extensive world knowledge and ability to flexibly incorporate user instructions~\cite{xi2024towards,zhang2023recommendation}. 
Their applications in RSs can be generally categorized into two paradigms. 
The first, LLM-based RSs, replace traditional recommendation models with either unmodified or specially adapted LLMs, yielding significant improvements, particularly in sequential recommendation tasks~\cite{bao2023bi,liao2024llara}. 
The second, LLM-enhanced RSs, leverage LLMs as representation generators for items, integrating them with traditional recommendation models, which has also led to substantial performance gains~\cite{ren2024representation,sheng2024language}.

Although the incorporation of LLMs has benefited RSs in multiple aspects, \textit{it has also introduced a severe challenge, i.e., unfairness}~\cite{zhang2023chatgpt,tommasel2024fairness}.
Regardless of the specific application paradigm, LLMs primarily operate on items, making Item-side Fairness (IF) a key concern.
\begin{figure}[t]
    \centering
    \includegraphics[width=0.95\linewidth]{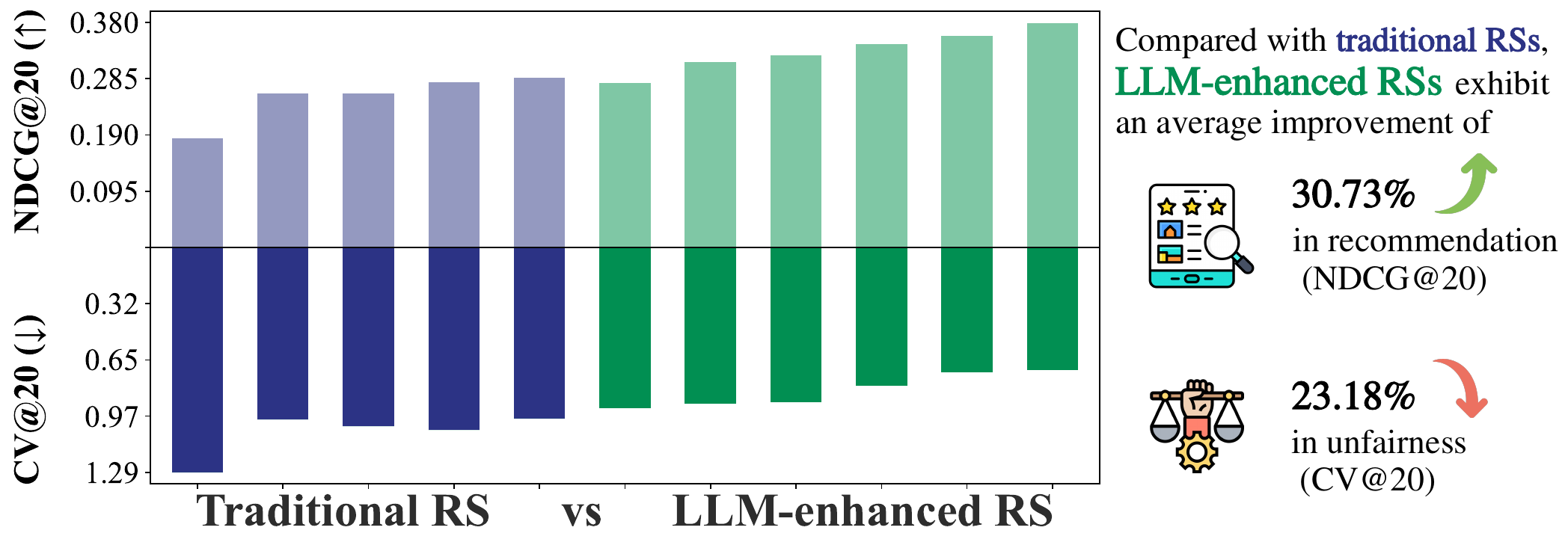}
    \caption{Average performance comparison of 5 traditional and 6 LLM-enhanced RSs with diverse architectures and base LLMs across three datasets. CV means coefficient of variation.}
    \label{fig:empi}
\end{figure}
IF requires that different item groups receive equal treatment~\cite{li2023fairness,wang2023survey}. 
Recent studies on IF in LLM-based RSs reveal that these models exhibit greater unfairness compared to traditional sequential RSs, uncovering additional fairness issues introduced by LLMss~\cite{jiang2024item,gao2024sprec}. Moreover, some studies have highlighted that LLMs inherently exhibit bias~\cite{yeh2023evaluating,bai2025explicitly}, including generating unfair outputs for different inputs~\cite{bai2024measuring}.

However, in LLM-enhanced RSs, whether similar fairness issues observed in LLM-based RSs emerge when LLMs are used to enhance item representations 
remains unexplored.
Thus, we first conduct an empirical study leveraging the latest LLM-enhanced RSs and various LLM backbones to compare the fairness differences between traditional methods and LLM-enhanced RSs (Section ~\ref{sec:empirical}).
Through our empirical study in Figure~\ref{fig:empi}, we find that rather than exacerbating unfairness, integrating LLMs in this scenario actually improves fairness across different popularity groups and genre groups.
The LLM-enhanced RS is a powerful recommendation backbone that inherently achieves better fairness.
Nevertheless, there remains substantial room for fairness improvement in LLM-enhanced RSs, as the bottom 25\% of groups perform at only one-third of the average level. This motivates further exploration into improving fairness in LLM-enhanced RSs.

However, the architectural differences and varying sources of unfairness between LLM-enhanced RSs and traditional recommendation models pose new challenges for fairness interventions.
In general, unfairness can be mitigated at three stages: pre-training, in-training, and post-training.
Specifically, pre-training methods primarily adjust data or embeddings~\cite{ekstrand2018all}, conventional in-training methods modify the model or loss function~\cite{burke2018balanced}, and post-training approaches only intervene in the final recommendation model or ranking~\cite{patro2022fair}. 
These methods all fail to simultaneously mitigate both the unfairness embedded in LLM-generated semantic representations and the unfairness arising during the recommendation model's training process.
Consequently, we decompose the fairness issue into two distinct components: i) \textbf{prior unfairness} originating from LLM-generated representations and ii) \textbf{training unfairness} arising from subsequent recommendation model training. Addressing fairness issues in LLM-enhanced RSs necessitates mitigating both types of unfairness simultaneously, a direction that has remained unexplored in prior works.

To mitigate both prior unfairness and training unfairness simultaneously, we propose \model{}, a fairness-aware training framework based on bi-level optimization. 
\model{} enhances fairness by updating two independent sets of learnable parameters: i) LLM-generated representations from item information and ii) trainable projector in the recommendation model. These parameters are optimized via a two-level nested optimization process. 
In the lower-level stage, we train the recommendation model weights by minimizing a fairness-aware recommendation loss, while keeping the learnable representations fixed. 
After training, we pass the optimal weights to the upper-level stage, where we minimize the same fairness-aware recommendation loss to update the representations. Through this iterative approach, we simultaneously mitigate both prior and training unfairness while preserving recommendation performance. 

Furthermore, existing fairness-aware loss methods used in RSs, e.g., Reweight~\cite{hu2008collaborative,jiang2024item} or Group Distributionally Robust Optimization (GroupDRO)~\cite{sagawadistributionally,wen2022distributionally}, neglect the dynamic training processes of RS and inter-group mutual dependency effects, leading to suboptimal fairness and heavily degraded recommendation performance.
Thus, we introduce an adaptive inter-group balancing mechanism for the fair recommendation loss. 
Specifically, we design a multi-objective optimization-inspired loss function that iteratively identifies the gradient directions benefiting all groups simultaneously. By prioritizing gradients aligned with multiple objectives over single group optimization, our designed loss function enables adaptive fairness calibration, explicitly integrating inter-group dependencies into RS training dynamics.

Our main contributions in this paper are highlighted as follows:
\begin{itemize}[leftmargin=*]\setlength{\itemsep}{-\itemsep}
    \item To the best of our knowledge, we are the first to systematically investigate and address the unfairness issue in LLM-enhanced RSs. We decompose unfairness into prior unfairness (originating from the LLM)
    and training unfairness (introduced during the subsequent recommendation training process). 
    \item Based on above formulation, we propose a bi-level optimization-based fairness-aware training framework that is applicable to any LLM-enhanced RS framework. Specifically, we train both LLM-generated representations and the recommendation projector to simultaneously mitigate prior unfairness and training unfairness.
    \item We further design a fairness-aware recommendation loss. Through an inter-group balancing mechanism, our method leverages gradient information from all groups to determine the final update direction, dynamically balancing group-wise performance.
    \item We conduct comprehensive experiments on multiple real-world datasets.
    Experimental results show that \model{} substantially improves different types of IF and is effective across various base LLMs and LLM-enhance RS architectures.
\end{itemize}

\section{PRELIMINARY}

\begin{table*}[t]
\caption{Performance comparison (traditional RSs vs LLM-enhanced RSs) across Accuracy, Pop. Fairness, and Genre Fairness metrics. Bold values indicate the best results, with improvements calculated relative to the best results of traditional RSs. Arrows ($\uparrow$/$\downarrow$) denote the preferred metric direction.}
\label{tab:emprical}
\centering
\resizebox{\textwidth}{!}{
\begin{tabular}{l|ccc|cc|cc}
    \toprule
    \multicolumn{1}{c|}{\multirow{2}{*}{Model}} & \multicolumn{3}{c|}{Accuracy} & \multicolumn{2}{c|}{Pop. Fairness} & \multicolumn{2}{c}{Genre Fairness} \\
& Recall$\uparrow$ & NDCG$\uparrow$  & HR$\uparrow$    & CV$\downarrow$ & MIN$\uparrow$   & CV$\downarrow$ & MIN$\uparrow$ \\
    \midrule
    MF    & 0.0713 & 0.1501 & 0.0655 & 0.7759 & 0.0154 & 1.6268 & 0.0097 \\
    MultVAE & 0.0963 & 0.2076 & 0.0821 & 0.6584 & 0.0314 & 1.2356 & 0.0176 \\
    LightGCN & 0.0991 & 0.2188 & 0.0870 & 0.7326 & 0.0291 & 1.1878 & 0.0185 \\
    SGL   & 0.1054 & 0.2312 & 0.0874 & 0.7372 & 0.0255 & 1.1601 & 0.0145 \\
    XSimGCL & 0.1096 & 0.2399 & 0.0940 & 0.8191 & 0.0189 & 1.1786 & 0.0140 \\
    \midrule
    Linear (Qwen) & 0.1106 & 0.2309 & 0.0976 & 0.5235 & 0.0500 & 1.0462 & 0.0309 \\
    Linear (Llama) & 0.1370 & 0.2832 & 0.1145 & 0.4006 & 0.0775 & 0.9575 & 0.0355 \\
    Linear (SFR) & 0.1601 & 0.3257 & 0.1339 & \textbf{0.3373}(-48.8\%) & \textbf{0.0991}(+215.6\%) & 0.8090 & 0.0494 \\
    AlphaRec (Qwen) & 0.1232 & 0.2630 & 0.1053 & 0.5887 & 0.0494 & 1.0073 & 0.0268 \\
    AlphaRec (Llama) & 0.1373 & 0.2872 & 0.1132 & 0.5026 & 0.0653 & 0.9075 & 0.0393 \\
    AlphaRec (SFR) & \textbf{0.1658}(+51.2\%) & \textbf{0.3397}(+41.6\%) & \textbf{0.1385}(+47.3\%) & 0.4364 & 0.0889 & \textbf{0.7996}(-55.1\%) & \textbf{0.0497}(+168.6\%) \\
    \bottomrule
    \end{tabular}%
}
\end{table*}
\subsection{LLM-enhanced Recommender Systems}
Incorporating LLMs into RSs facilitates the derivation of item representations from rich textual metadata, rather than relying solely on discrete identifiers.
LLM-enhanced RSs are commonly structured in a two-stage paradigm: stage I employs LLMs to generate semantic representations, while stage II utilizes a trainable projector to refine these representations for downstream recommendation tasks.

Specifically, given a user-item interaction dataset $ \mathcal{D} = \{(u, i, y_{u,i}) \mid u \in U, i \in I\} $, where $ U \in \mathbb{R}^{|U|} $ denotes the set of users, $ I \in \mathbb{R}^{|I|} $ denotes the set of items, and $ y_{u,i} \in \{0,1\} $ indicates whether user $ u $ has interacted with item $ i $.  
\textbf{In stage I}, for an item $ i $ with textual description $ x_i $, a frozen LLM encodes it into a semantic vector representation $ z_i = f_{\mathrm{LLM}}(x_i) $, where $ f_{\mathrm{LLM}} $ is typically a pre-trained LLM or LLM-finetuned embedding model. User representations are often obtained by aggregating item representations, i.e., $z_u = \frac{1}{|I_u|} \sum_{i \in I_u} z_i$ , where $I_u$ denotes the set of items that user $u$ has interacted with. 
\textbf{In stage II}, to align the semantic  representation space to the recommendation space, a trainable projector with parameters $\boldsymbol{\theta}$ transforms $ z_i $ into the final item representation $ e_i = g_{\boldsymbol{\theta}}(z_i) $. Similarly, the final user representation is obtained as $ e_u = g_{\boldsymbol{\theta}}(z_u) $, ensuring that both user and item representations exist in the same latent space for effective recommendation. The likelihood of user $ u $ interacting with item $ i $ is then estimated via a scoring function $ \hat{y}_{u,i} = s(e_u, e_i) $, where $ s $ quantifies the relevance between user and item representations.

The training objective for the recommendation model is formulated as $ \boldsymbol{\theta}^* = \mathop{\arg\min}\limits_{\boldsymbol{\theta}} L_{\mathrm{rec}}(\mathcal{D}) $, where $ L_{\mathrm{rec}} $ represents the recommendation loss, such as BPR loss~\cite{rendle2009bpr} or InfoNCE loss~\cite{oord2018representation}.

\subsection{IF Definition in Recommender Systems}  

Fairness in RSs extends beyond users to ensure that items from different groups achieve comparable recommendation performance. Given a set of $N$ item groups $G_I=\{G_1,G_2,...,G_N\}$, these groups can be defined based on various criteria, such as item popularity or specific attributes (e.g., genre for movies, category for products, or source for news). 

To assess fairness across item groups, we define the utility of each group as a measure of the recommendation performance of potential interactions within that group.
Specifically, let $U$ denote the set of all users. Following ~\cite{chen2023fairly,wang2024intersectional}, the utility of an item group $G_n$ is computed as the average utility of its items across all relevant users:
\begin{equation}
\label{define}
    \text{G\_Utility}(n)@K = \frac{1}{|U|} \sum_{u \in U} \mathrm{utility}(u, n)@K,
\end{equation}  
where $\mathrm{utility}(u, n)@K$ quantifies the recommendation performance of items in $G_n$ for user $u$. This utility can be measured using various ranking-based metrics, such as Recall, Hit Rate (HR), or Normalized Discounted Cumulative Gain (NDCG). 
Built upon this utility definition, IF ensures that different item groups achieve comparable utility values, thereby promoting an equitable distribution of recommendation quality across groups. 
Consistent with \cite{wang2024intersectional}, we define $\epsilon$-Item-side Fairness as follows:  

\textbf{Definition 1 ($\epsilon$-Item-side Fairness).} A RS satisfies $\epsilon$-item-side fairness if, for any two item groups $G_a$ and $G_b$, the difference in their utilities is bounded by a predefined threshold $\epsilon$: $| \mathrm{G\_Utility}(a) - \mathrm{G\_Utility}(b) | \leq \epsilon$, where $\epsilon$ is a sufficiently small positive constant.  


\section{THE BIFAIR FRAMEWORK}
\begin{figure*}[t]
    \centering
    \includegraphics[width=\linewidth, trim=2cm 4.6cm 2cm 4cm, clip]{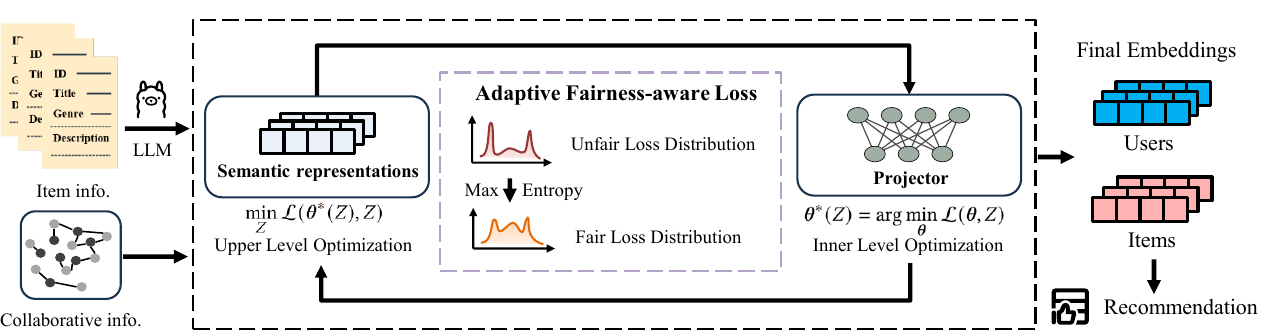}
    \caption{An illustration of the pipeline of our proposed \model{}. The representations generated by LLMs and the trainable projector are jointly optimized within a bi-level framework, where an adaptive fairness-aware recommendation loss is employed to balance group-wise performance. \model{} simultaneously mitigates both prior unfairness and training unfairness.}
    \label{fig:frame}
\end{figure*}

In this section, we first conduct an empirical study on IF in LLM-enhanced RSs. Our analysis reveals that, compared to traditional RSs, LLM-enhanced RSs exhibit improved fairness. 
However, there remains significant room for improvement, as the performance on certain challenging groups continues to lag behind optimal levels.
To further enhance fairness, 
we introduce \model{}, a fairness-aware training framework for LLM-enhanced RSs to ensure fairness among item groups. As illustrated in Figure~\ref{fig:frame}, our approach consists of two key components: a bi-level optimization framework and an adaptive fairness-aware recommendation loss.

\subsection{Empirical Study}
\label{sec:empirical}
To investigate fairness issues in LLM-enhanced RSs, we utilize three categories from the Amazon Review dataset.
Specifically, we compare the fairness performance of traditional RSs with that of LLM-enhanced RSs, including Linear Mapping (Linear) and AlphaRec~\cite{sheng2024language}. Table~\ref{tab:emprical} presents the results on the Book dataset.
Further details regarding the datasets, models, evaluation metrics, and comprehensive experimental results are provided in Section~\ref{section:setting} and Appendix~\ref{app:exp}.

We find that while Linear and AlphaRec significantly enhance overall model performance, they also substantially improve fairness compared to traditional ID-based RSs. Regardless of the fairness metric, the performance improvement of the worst-performing group (MIN) far exceeds the average accuracy gain, as LLMs provide stronger representation initialization for disadvantaged groups. Additionally, LLM-finetuned embedding model (e.g., SFR) demonstrate greater fairness than standard LLM.

However, even the best-performing LLM-enhanced RS in terms of Genre Fairness demonstrates that the performance on the worst-performing groups remains \textbf{less than one-third} of the overall average.
Moreover, more complex model architectures (e.g., AlphaRec) tend to exacerbate Pop. Unfairness compared to simpler architectures (e.g., Linear).

\subsection{Bi-Level Optimization Framework}

LLM-enhanced RSs follow a two-stage paradigm, where LLMs are used to generate representations, followed by training a projector. Each stage introduces distinct sources of unfairness: the LLMs inherently carry biases stemming from their pre-training and fine-tuning corpora (stage I), while the projector training is exposed to popular bias prevalent in recommendation data (stage II).
To mitigate the unfairness in both stage, we first decompose the unfairness issue into two distinct components: i) \textbf{prior unfairness} originating from LLM-generated representations and ii) \textbf{training unfairness} arising from subsequent recommendation model training. We address these two sources of unfairness simultaneously via a bi-level optimization framework.

\subsubsection{Inner-Level Optimization}
To mitigate the \textit{training unfairness}, we optimize the trainable projector \( g_{\boldsymbol{\theta}} \), ensuring balanced learning across different item groups:  
\begin{equation}
    \boldsymbol{\theta}^*(Z) = \arg\min_{\boldsymbol{\theta}} \mathcal{L}(\boldsymbol{\theta}, Z),
\end{equation}
where \( Z \) is the semantic representation generated by LLM and $\mathcal{L}(\cdot)$ is a fairness-aware recommendation loss to ensure 
fair learning across different item groups. 
This optimization mitigates the risk of disproportionately favoring certain item groups due to variations in representation quality or data distribution, thereby promoting fair training dynamics.  

\subsubsection{Outer-Level Optimization}
To address the \textit{prior unfairness} inherited from the pre-trained LLM, we optimize the semantic representation \( Z \) by minimizing the same fairness-aware loss:  
\begin{equation}
    \min_{Z} \mathcal{L}(\boldsymbol{\theta}^*(Z), Z).
\end{equation}
Since the LLM remains frozen, its representations may inherently encode biases originating from pre-training data, leading to uneven representation quality across different item groups. Optimizing \( Z \) allows us to refine these representations within the recommendation space, mitigating biases and ensuring a more fair distribution for future training.  

\subsubsection{Bi-Level Optimization Framework} 
Combining the above two levels, we establish the following bi-level optimization framework:  
\begin{align}
    & \min_{Z} \mathcal{L}(\boldsymbol{\theta}^*(Z), Z) \\ 
    & \mathrm{s.t.} \quad \boldsymbol{\theta}^*(Z) = \arg\min_{\boldsymbol{\theta}} \mathcal{L}(\boldsymbol{\theta}, Z).\nonumber
\end{align}
The inner-level optimization guarantees fairness during model training, while the outer-level optimization refines the semantic space to mitigate pre-existing biases from the LLM. Together, they form a fairness-aware recommendation framework that tackles both training and prior unfairness simultaneously.  

\subsubsection{Solving the Bi-Level Optimization Problem}
Directly computing the gradient of the outer-level objective is computationally expensive due to the costly inner optimization. To alleviate this, we adopt an approximate gradient scheme inspired by ~\cite{liudarts} as follows:  
\begin{align}
    \nabla_Z \mathcal{L}(\boldsymbol{\theta}^*(Z), Z) \approx \nabla_Z \mathcal{L}(\boldsymbol{\theta} - \xi \nabla_{\boldsymbol{\theta}} \mathcal{L}(\boldsymbol{\theta}, Z), Z),
\end{align}
where \( \boldsymbol{\theta} \) represents the current trainable projector parameters, and \( \xi \) is a small step size for approximating the inner-level update. Instead of fully solving for \( \boldsymbol{\theta}^*(Z) \), we approximate it using only a single step of gradient descent, significantly reducing computational cost. 

\begin{algorithm}[tb]
\caption{Bi-Level Optimization Framework}
\label{alg:bilevel}
\textbf{Input}: Dataset $\mathcal{D}$, Learning Rate $\eta$, Step Size $\xi$, Regularization $\epsilon$.\\

\textbf{Output}: Optimized Model Parameters $\boldsymbol{\theta}_T$, Refined Representations $Z_T$.

\begin{algorithmic}[1]
\STATE Initialize model parameters $\boldsymbol{\theta}_0$ and generate semantic representations $Z_0$.
\FOR{$t$ in $1 \dots T$}
    \STATE \textbf{Inner-Level Optimization:} 
    \STATE \quad $\boldsymbol{\theta}_t \gets \boldsymbol{\theta}_{t-1} - \eta \nabla_{\boldsymbol{\theta}} \mathcal{L}(\boldsymbol{\theta}, Z_{t-1})$.
    \STATE \textbf{Outer-Level Optimization:}  
    \STATE \quad Compute approximate gradient: 
    \STATE \quad $\nabla_Z \mathcal{L}(\boldsymbol{\theta}', Z) - \xi \nabla^2_{Z, \boldsymbol{\theta}} \mathcal{L}(\boldsymbol{\theta}, Z) \nabla_{\boldsymbol{\theta}'} \mathcal{L}(\boldsymbol{\theta}', Z)$.
    \STATE \quad Approximate second-order term using finite differences:
    \STATE \quad $\frac{\nabla_Z \mathcal{L}(\boldsymbol{\theta}^+, Z) - \nabla_Z \mathcal{L}(\boldsymbol{\theta}^-, Z)}{2\epsilon}$.
    \STATE \quad $Z_t \gets Z_{t-1} - \eta \nabla_Z \mathcal{L}(\boldsymbol{\theta}_t, Z_{t-1})$.
\ENDFOR
\STATE \textbf{return} Optimized Model $\boldsymbol{\theta}_T$, Refined Representations $Z_T$.
\end{algorithmic}
\end{algorithm}

Applying the chain rule to this approximation yields:  
\begin{align}
    \nabla_Z \mathcal{L}(\boldsymbol{\theta}', Z) - \xi \nabla^2_{Z, \boldsymbol{\theta}} \mathcal{L}(\boldsymbol{\theta}, Z) \nabla_{\boldsymbol{\theta}'} \mathcal{L}(\boldsymbol{\theta}', Z),
\end{align}
where \( \boldsymbol{\theta}' = \boldsymbol{\theta} - \xi \nabla_{\boldsymbol{\theta}} \mathcal{L}(\boldsymbol{\theta}, Z) \) are the parameters after a single gradient update. Direct computation of the second-order derivative term is computationally prohibitive; however, we employ a finite difference approximation to alleviate this complexity. Specifically, given a small scalar \( \epsilon \), we can approximate:  
\begin{align}
    \nabla^2_{Z, \boldsymbol{\theta}} \mathcal{L}(\boldsymbol{\theta}, Z) \nabla_{\boldsymbol{\theta}'} \mathcal{L}(\boldsymbol{\theta}', Z) \approx \frac{\nabla_Z \mathcal{L}(\boldsymbol{\theta}^+, Z) - \nabla_Z \mathcal{L}(\boldsymbol{\theta}^-, Z)}{2\epsilon},
\end{align}
where \( \boldsymbol{\theta}^{\pm} = \boldsymbol{\theta} \pm \epsilon \nabla_{\boldsymbol{\theta}'} \mathcal{L}(\boldsymbol{\theta}', Z) \). This reduces computational complexity from \( O(|Z| \times |\boldsymbol{\theta}|) \) to \( O(|Z| + |\boldsymbol{\theta}|) \), making the optimization significantly more efficient. The entire optimization process is summarized in Algorithm \ref{alg:bilevel}. 

\subsection{Adaptive Fairness-aware Recommendation Loss}
In the training process of RSs, it is crucial to ensure that the model learns from different item groups in a balanced manner, avoiding overfitting to specific groups. 
In the context of LLM-enhanced RSs, achieving such balance becomes even more challenging, as it must be accomplished without relying on prior information about the groups. 
Motivated by principles from multi-objective optimization~\cite{desideri2012multiple}, we propose an adaptive fairness-aware recommendation loss. 
This loss function ensures that the model does not disproportionately favor any particular group during optimization but instead dynamically adjusts to account for the learning process across all groups. 
Specifically, we define a fairness-aware \( N \)-dimensional group loss vector as follows:
\begin{equation}
    \mathcal{L}(\boldsymbol{\theta}) = (\mathbb{E}_{(u, i) \sim P_{G_1}}[\ell(\boldsymbol{\theta})], \dots, \mathbb{E}_{(u, i) \sim P_{G_N}}[\ell(\boldsymbol{\theta})]),
\end{equation}
where $P_{G_n}(U, I) = P(U) P(I \mid G_n)$, $(u,i) \sim P_{G_n}$ indicates that the training samples are drawn from the data distribution of item group $G_n$, $\ell(\cdot)$ represents the recommendation loss, and $f_{\boldsymbol{\theta}}$ denotes the entire recommendation model. 
Unlike prior fairness-aware approaches that rely on prior weights $w$ derived from the data distribution~\cite{zhao2023popularity,jiang2024item}, we aim to adaptively learn the group weights $w$ during optimization to ensure balanced learning process across different groups, thereby achieving fairness in training.

To achieve this, we introduce the entropy function \( H(\cdot) \), which measures the diversity of a continuous distribution. 
For clarity, we define $H_s(\cdot)$ as the entropy computed after applying the softmax function.
We formalize fairness in training as the maximization of entropy \( H_s(\mathcal{L}(\boldsymbol{\theta})) \) with respect to the loss distribution. 
The optimal fairness-aware weight $ w^* $ is obtained by solving:
\begin{align}
    w^* &= \arg\max_{w \in \Delta_N} H_s(\mathcal{L}(\boldsymbol{\theta}^*(w))) \\
    & \mathrm{s.t.} \quad \boldsymbol{\theta}^*(w) = \arg\min_{\boldsymbol{\theta}} \sum_{n=1}^{N} w_n \ell_n(\boldsymbol{\theta}_t), \nonumber
\end{align}
where $\ell_n(\boldsymbol{\theta}) = \mathbb{E}_{(u, i) \sim P_{G_n}}[\ell(\boldsymbol{\theta})]$.
Intuitively, $ w^* $ is obtained by maximizing $ H_s(\mathcal{L}(\boldsymbol{\theta})) $ while ensuring that it does not conflict with the gradient vectors of each group.

As a result, we reformulate the fairness-aware weight optimization problem as:
\begin{align}
\label{equ:w}
    w^* &= \arg\max_{w \in \Delta_N} d_t(w)^\top \sum_{n=1}^{N} \nabla \ell_n(\boldsymbol{\theta}_t) \left[ p_n \log p_n - p_n \sum_{i=1}^{N} p_i \log p_i \right] \\
    & \mathrm{s.t.} \quad d_t(w)^\top \nabla \ell_n(\boldsymbol{\theta}_t) \geq 0 \quad \forall n \in [N], \nonumber
\end{align}
where $ p_n = \frac{e^{\ell_n(\boldsymbol{\theta}_t)}}{\sum_{i=1}^{N} e^{\ell_i(\boldsymbol{\theta}_t)}} $ and \( d_t(w) \) represents the gradient-weighted update vector under the current weight assignment \( w \). The detailed derivation of Eq.~(\ref{equ:w}) is provided in Appendix~\ref{app:derivation}. This constrained optimization problem can be interpreted as searching for an update direction that maximizes entropy-driven fairness while satisfying per-group descent constraints.

However, strictly adhering to the update direction that minimizes the training error of high-loss groups may yield suboptimal outcomes, as these groups could contain noisy data or biases introduced by LLMs. Furthermore, certain low-loss groups may benefit from superior representation projections from the language space to the recommendation space. To preserve overall recommendation quality, we impose a minimum guarantee on the solutions for low-loss groups, ensuring that their updates are at least aligned with the objective of maximizing \( H_s(\vec{\mathcal{L}}(\boldsymbol{\theta})) \). 

For implementation, we adopt the Frank-Wolfe algorithm~\cite{jaggi2013revisiting}, a well-established solver for minimum-norm point problems, to find a common descent direction that balances the contributions of all gradient vectors, including the entropy component. The entropy-regularized fair weighting via Frank-Wolfe is presented in Algorithm~\ref{alg:fw-entropy}.

Finally, we compute the parameter update direction based on the optimal weight assignment \( w^* \):
\begin{equation}
    d_t(w^*) = \sum_{n=1}^{N} w_n^* \nabla \ell_n(\boldsymbol{\theta}_t),
\end{equation}
and update the model parameters as follows: $ \boldsymbol{\theta}_{t+1} = \boldsymbol{\theta}_t - \eta d_t(w^*)$, where \( \eta \in \mathbb{R}^+ \) is the learning rate. 

This fairness-aware optimization framework ensures that the RS considers the learning process of all item groups throughout training while dynamically adjusting the group weights via entropy maximization.
By doing so, our approach prevents the optimization process from disproportionately focusing on any specific group.
Furthermore, our entropy-based approach not only fosters fairness across all groups but also improves the stability of the optimization process.
\begin{algorithm}[tb]
\caption{Entropy-Regularized Fair Weighting via Frank-Wolfe}
\label{alg:fw-entropy}
\textbf{Input}: Gradient set $\mathcal{G} = \{\nabla \ell_1, \ldots, \nabla \ell_N, -\nabla H_s(\mathcal{L})\}$, number of iterations $T$.

\textbf{Output}: Optimal fairness-aware weight $w_T$.

\begin{algorithmic}[1]
\STATE Initialize $w_0 \gets \left[ \frac{1}{|\mathcal{G}|}, \ldots, \frac{1}{|\mathcal{G}|} \right]$.
\STATE Form Gram matrix $B \gets \mathcal{G}^\top \mathcal{G}$.
\FOR{$t = 0$ \textbf{to} $T-1$}
    \STATE $v_t \gets \arg\min_{v \in \{\boldsymbol{e}_1, \ldots, \boldsymbol{e}_{|\mathcal{G}|}\}} v^\top B w_t$.
    \STATE $\gamma_t \gets \arg\min_{\gamma \in [0, 1]} \left\| (1 - \gamma) w_t + \gamma v_t \right\|_B^2$.
    \STATE $w_{t+1} \gets (1 - \gamma_t) w_t + \gamma_t v_t$.
\ENDFOR
\STATE \textbf{return} $w_T$.
\end{algorithmic}
\end{algorithm}

Importantly, we compute the fairness-aware weight $w$ in Eq.~(\ref{equ:w}) only after determining the update directions for both the inner and outer levels. This approach significantly reduces computational overhead compared to a naive method that would require recalculating $w$ at every gradient step. Without this simplification, the optimization process would resemble a tri-level problem, introducing prohibitive complexity due to nested dependencies. By deferring the computation of fairness-aware weights until the update direction is finalized, we achieve computational efficiency while ensuring balanced learning across item groups.



\section{EXPERIMENTS}
\label{section:setting}

\subsection{Datastes}
We conduct experiments on three real-world datasets from the Amazon platform\footnote{http://jmcauley.ucsd.edu/data/amazon/} that are widely used in recommendation tasks. The statistical details of datasets are summarized in Appendix ~\ref{app:dataset}. \textbf{Movies}: provides user reviews and detailed metadata for movies and TV products. \textbf{Games}: contains user reviews and metadata for video games and related products. \textbf{Books}: covers user reviews and metadata for a wide range of book products.

We adopt the pre-processing procedure from previous studies~\cite{zhang2024generative,yang2023towards}. We considering each user interaction with a rating of no less than 3 as a valid interaction. The entire historical interaction data is partitioned into training, validation, and test sets in a 4:3:3 ratio. To mitigate the cold-start problem, we exclude users with fewer than 20 interactions and remove items from the validation and test sets that do not appear in the training set.

To evaluate group fairness, we implement two item grouping strategies : based on popularity and genre. For Popularity Fairness (Pop. Fairness), we adopt an approach wherein items are ranked in accordance with their respective interaction counts. Following the prior works ~\cite{li2021user,han2023processing}, we delineate the top 10\% of items as one group, and the remaining 90\% as another group. For Genre Fairness, we assign items to groups directly based on the genres extracted from the dataset. 
\subsection{Base Models}
For traditional RSs, we select MF~\cite{koren2009matrix}, MultVAE~\cite{liang2018variational}, LightGCN~\cite{he2020lightgcn}, SGL~\cite{wu2021self}, and XSimGCL~\cite{yu2023xsimgcl}. For LLM-enhanced RSs, we choose Linear and AlphaRec~\cite{sheng2024language} as representative models. Linear and AlphaRec follow a typical paradigm where an LLM generates presentations, which are then mapped from the semantic space to the recommendation representation space using a relatively simple projector. As a standard framework for LLM-enhanced RSs, AlphaRec serves as a foundation upon which existing and future approaches can be regarded as extensions. For the base LLMs, we select Qwen2.5-7B (Qwen)~\cite{yang2024qwen2}, Llama2-7B (Llama)~\cite{touvron2023llama}, and SFR-Embedding-Mistral-7B (SFR)~\cite{meng2024sfrembedding}.
\subsection{Experimental Settings}
\subsubsection{Compared Methods}
We compare \model{} with several representative fair training frameworks, including:
\begin{itemize}[leftmargin=*]\setlength{\itemsep}{2pt}
    \item \textbf{MultiFR}~\cite{wu2022multi} formulates recommendation loss and fairness constraints as two distinct tasks and employs MGDA to identify a solution with the minimum Euclidean norm over the composite gradient.
    \item \textbf{Reweight}~\cite{hu2008collaborative} is a simple yet robust approach applicable to both conventional RSs and LLM-based RSs~\cite{jiang2024item}. It typically assigns weights based on the inverse of group frequency.
    \item \textbf{PopDRO}~\cite{zhao2023popularity} incorporates item popularity into the DRO formulation to improve recommendation performance.
    \item \textbf{DRORec}~\cite{wen2022distributionally} introduces DRO into recommender systems and enhances its effectiveness by leveraging uncertainty-aware subgrouping and streaming estimations.
    \item \textbf{ITFR}~\cite{wang2024intersectional} integrates sharpness-aware disadvantage group discovery with GroupDRO to address both user-side and item-side unfairness.
\end{itemize}
\subsubsection{Evaluation Metrics}
To evaluate recommendation performance, we employ three widely used metrics, i.e., Recall ($\mathrm{Recall}@K$), Normalized Discounted Cumulative Gain ($\mathrm{NDCG}@K$) and Hit Ratio ($\mathrm{HR@}K$), to evaluate the recommendation performance of the LRS on the testing set.

To quantify IF, given the utility definition in Eq. (\ref{define}), let \( \mathrm{Utility} \) denote the set of utilities for all item groups. We employ the coefficient of variation (CV) as a measure of utility dispersion among item groups~\cite{chen2023fairly}: $\mathrm{CV}@K = \frac{\operatorname{std}(\mathrm{Utility})}{\operatorname{mean}(\mathrm{Utility})}$. A lower $\mathrm{CV}@K$ value indicates better item-side fairness.  

In addition to measuring overall dispersion, we also introduce a worst-case fairness metric $\mathrm{MIN}@K$ to capture extreme utility imbalances. Specifically, instead of focusing on a single worst-performing group, which may be unstable ~\cite{li2021leave}. In reference to previous research~\cite{wang2024intersectional}, we measure the average utility of the bottom 25\% of item groups. We set $K=20$.
\begin{table*}[t]
\caption{Performance comparison across Accuracy, Pop. Fairness, and Genre Fairness metrics. Bold values indicate the best results. Arrows ($\uparrow$/$\downarrow$) denote the preferred metric direction.}
\label{tab:main}
\centering
\resizebox{\textwidth}{!}{
\begin{tabular}{lccccc|ccccc|ccccc}
    \toprule
    Dataset & \multicolumn{5}{c|}{Movies}           & \multicolumn{5}{c|}{Games}            & \multicolumn{5}{c}{Books} \\
    \midrule
    IF type & \multicolumn{15}{c}{Pop. Fairness} \\
    \midrule
     & Recall$\uparrow$ & NDCG$\uparrow$  & HR$\uparrow$    & CV$\downarrow$ & MIN$\uparrow$   & Recall$\uparrow$ & NDCG$\uparrow$  & HR$\uparrow$    & CV$\downarrow$ & MIN$\uparrow$   & Recall$\uparrow$ & NDCG$\uparrow$  & HR$\uparrow$    & CV$\downarrow$ & MIN$\uparrow$ \\
    \midrule
    AlphaRec & \textbf{0.2034} & \textbf{0.1831} & 0.4338 & 0.3653 & 0.1089 & \textbf{0.1750} & 0.1250 & 0.3661 & 0.4388 & 0.0894 & \textbf{0.1658} & 0.1385 & 0.3397 & 0.4364 & 0.0889 \\
    MultiFR & 0.2004 & 0.1768 & 0.4278 & 0.4001 & 0.1003 & 0.1707 & 0.1211 & 0.3601 & 0.5238 & 0.0732 & 0.1603 & 0.1322 & 0.3299 & 0.5820 & 0.0638 \\
    RW    & 0.2017 & 0.1783 & 0.4313 & 0.3223 & 0.1159 & 0.1706 & 0.1208 & 0.3570 & 0.4095 & 0.0918 & 0.1611 & 0.1327 & 0.3318 & 0.6064 & 0.0607 \\
    PopDRO  & 0.1950 & 0.1711 & 0.4212 & 0.3413 & 0.1087 & 0.1695 & 0.1187 & 0.3523 & 0.4321 & 0.0911 & 0.1658 & 0.1380 & 0.3390 & 0.4670 & 0.0839 \\
    DRORec & 0.2025 & 0.1801 & 0.4328 & 0.3173 & 0.1100 & 0.1727 & 0.1230 & 0.3613 & 0.3908 & 0.0929 & 0.1656 & \textbf{0.1388} & \textbf{0.3411} & 0.4115 & 0.0902 \\
    ITFR  & 0.2028 & 0.1818 & \textbf{0.4341} & 0.3065 & 0.1197 & 0.1723 & 0.1224 & 0.3602 & 0.3725 & 0.0935 & 0.1622 & 0.1335 & 0.3329 & 0.4098 & 0.0917 \\
    \model{} & 0.2029 & 0.1820 & 0.4305 & \textbf{0.2860} & \textbf{0.1232} & 0.1721 & \textbf{0.1261} & \textbf{0.3679} & \textbf{0.3451} & \textbf{0.1039} & 0.1651 & 0.1374 & 0.3382 & \textbf{0.3560} & \textbf{0.0944} \\
    \midrule
    IF type & \multicolumn{15}{c}{Genre Fairness} \\
    \midrule
     & Recall$\uparrow$ & NDCG$\uparrow$  & HR$\uparrow$    & CV$\downarrow$ & MIN$\uparrow$   & Recall$\uparrow$ & NDCG$\uparrow$  & HR$\uparrow$    & CV$\downarrow$ & MIN$\uparrow$   & Recall$\uparrow$ & NDCG$\uparrow$  & HR$\uparrow$    & CV$\downarrow$ & MIN$\uparrow$ \\
    \midrule
    AlphaRec & 0.2034 & \textbf{0.1831} & 0.4338 & 0.9138 & 0.0313 & 0.1750 & 0.1250 & 0.3661 & 0.3896 & 0.0788 & \textbf{0.1658} & \textbf{0.1385} & \textbf{0.3397} & 0.8090 & 0.0497 \\
    MultiFR & 0.1964 & 0.1690 & 0.4204 & 0.9827 & 0.0184 & 0.1501 & 0.1059 & 0.3191 & 0.4734 & 0.0515 & 0.1485 & 0.1206 & 0.3025 & 0.8712 & 0.0418 \\
    RW    & 0.1965 & 0.1719 & 0.4187 & 0.8895 & 0.0321 & 0.1673 & 0.1159 & 0.3501 & 0.4638 & 0.0387 & 0.1632 & 0.1342 & 0.3327 & 0.77  & 0.0603 \\
    PopDRO  & 0.1921 & 0.1673 & 0.4102 & 0.8815 & 0.0357 & 0.1732 & 0.1229 & 0.3632 & 0.3998 & 0.0703 & 0.1649 & 0.1363 & 0.3371 & 0.7934 & 0.054 \\
    DRORec & 0.2032 & 0.1821 & 0.4337 & 0.9307 & 0.0261 & 0.1707 & 0.1208 & 0.3582 & 0.3663 & 0.0874 & 0.1603 & 0.1335 & 0.3241 & 0.7886 & 0.0552 \\
    ITFR  & \textbf{0.2035} & 0.1818 & \textbf{0.4350} & 0.8688 & 0.0369 & 0.1755 & 0.1253 & 0.3698 & 0.3605 & 0.0873 & 0.1652 & 0.1369 & 0.3393 & 0.775 & 0.0596 \\
    \model{} & 0.2016 & 0.1804 & 0.4322 & \textbf{0.8437} & \textbf{0.0382} & \textbf{0.1775} & \textbf{0.1258} & \textbf{0.3705} & \textbf{0.3018} & \textbf{0.0955} & 0.1654 & 0.1355 & 0.3392 & \textbf{0.7122} & \textbf{0.0648} \\
    \bottomrule
    \end{tabular}%
}
\end{table*}
\subsubsection{Implement Details}
\begin{itemize}[leftmargin=*]\setlength{\itemsep}{2pt}
    \item \textbf{Recommendation Models:} For all traditional RSs, we set the number of graph propagation layers to 2 and the embedding dimension to 64 by default. For AlphaRec, unless otherwise specified, the base LLM is set to SFR, which achieves the best recommendation performance.
    \item \textbf{Training Settings:} All fairness methods are implemented on the corresponding LLM-enhanced RS architectures. All models are optimized using the Adam optimizer. The number of negative samples is set to 256. We apply early stopping with a maximum of 500 training epochs. Training is terminated if the $\mathrm{Recall}@20$ on the validation set does not improve for 20 consecutive epochs. The all ranking evaluation protocol~\cite{krichene2020sampled} is adopted for all experiments. The base recommendation loss is set to the InfoNCE loss~\cite{oord2018representation}.
    \item \textbf{Hyper-parameters:} Following prior work ~\cite{sheng2024language}, we set the hyperparameter $\tau$ of AlphaRec to 0.15 for the Movies and Books datasets, and 0.2 for Games. For the bi-level optimization setup, we set the $\beta$ of the AdamW optimizer to (0.9, 0.999) and apply a weight decay of 0.1. The learning rate decreases progressively following a polynomial decay with exponent 0.9 consistent with previous studies ~\cite{zhang2023customized,zhang2024blo}, based on the ratio of current to maximum training iterations.
    \item \textbf{Hardware Information:} All experiments are conducted on the same Ubuntu 20.04 LTS server equipped with a 48-core CPU, 256GB RAM, and an NVIDIA A800 GPU. 
\end{itemize}
More experimental details are provided in Appendix~\ref{app:exp}.

\subsection{Results and Analysis}
\subsubsection{Fairness Performance}
As reported in Table ~\ref{tab:main}, across all datasets, \model{} achieves the best fairness performance. Compared to the strongest baseline, \model{} reduces the CV by an average of 9.05\% and improves the MIN score by an average of 5.63\% in the Pop. Fairness setting. Thus, \model{} reduces popular bias and increasing recommendation performance for underrepresented items.

In terms of Genre Fairness, \model{} further reduces CV by average 8.88\% and improves MIN scores by average 6.74\% across all the datasets. Notably, on Games, \model{} achieves the largest reduction in CV (16.28\%). These results confirm that \model{} effectively maintain group-level utility balance. 

\textbf{Summary}: \model{} achieves significant fairness improvements across all datasets, with enhanced group fairness over competitive baselines.
\subsubsection{Recommendation Performance}
We also evaluate the effect of \model{} on recommendation performance. in Table ~\ref{tab:main}, even when compared with strong accuracy-focused baseline AlphaRec, \model{} achieves comparable or slightly better results, especially on Games dataset. This indicates that the fairness optimization in \model{} does not compromise utility. Compared to all fair training frameworks, \model{} consistently maintains upstream-level recommendation performance while substantially surpassing them in fairness.

\textbf{Summary}: \model{} delivers fairness gains without sacrificing accuracy, outperforming or matching state-of-the-art baselines.
\subsubsection{Performance on different Base LLMs and RS Structures}
We further assess the generalizability of \model{} across different LLM backbones and recommendation architectures. In Table ~\ref{tab:lrs}, for Linear that already exhibit better Pop. Fairness, \model{} still brings significant fairness improvements of 2.44\% in terms of the CV metric, highlighting its robustness. Morevoer, as reported in Table ~\ref{tab:llm}, when applied to other generic LLMs or LLM-based fine-tuned embedding models, \model{} continues to deliver substantial fairness gains. Unlike methods that rely on architecture-specific design, \model{} offers robustness under varying configurations. It generalizes well across both original and embedding LLMs, suggesting broad applicability in real-world RSs.

\textbf{Summary}: \model{} generalizes across diverse model backbones and architectures, achieving consistent fairness gains.
\subsubsection{In-depth Analysis} We observe that many of the compared methods are directly effective in LLM-enhanced RSs. However, their performance still lags behind \model{}, and we now analyze the reasons behind this gap.

The suboptimal performance of MultiFR can be attributed to direct adoption of MGDA, which tends to favor objectives with smaller gradient magnitudes~\cite{liutowards}, thereby undermining fairness guarantees in multi-group scenarios. The Reweight method relies on the estimation of fairness-aware weights; however, the presence of prior unfairness severely hinders accurate estimation. The remaining three baselines inherit the core idea of GroupDRO, but overlook the collaborative relationships among groups, which are essential for improving overall performance. As a result, they struggle to enhance the performance of the worst-performing groups, limiting their ability to improve fairness while achieving stable and high-quality recommendation performance. Furthermore, all compared methods solely address training unfairness and are incapable of mitigating prior unfairness, leading to suboptimal fairness outcomes. 

In contrast, \model{} adopts a bi-level optimization framework to simultaneously tackle both types of unfairness. And the introduced adaptive fairness-aware loss effectively balances performance across different groups. By adopting a multi-task optimization formulation and entropy regularization, it further promotes knowledge transfer among groups. This design not only enhances the overall performance but also significantly improves the outcomes of the worst-performing group.

\subsection{Ablation Studies}
\begin{figure}
    \centering
    \includegraphics[width=\linewidth]{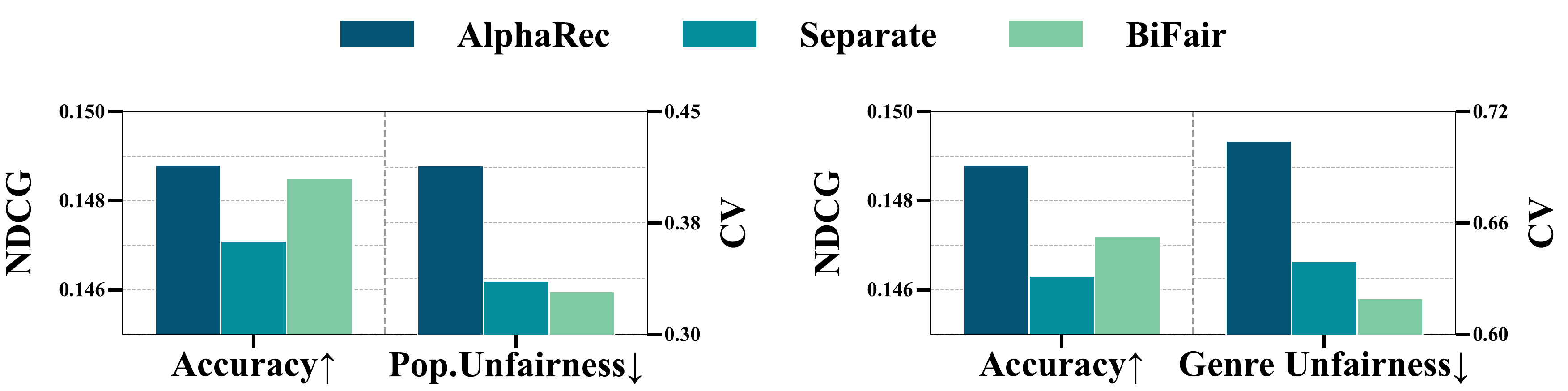}
    \caption{Ablation study of bi-level optimization framework with separate training. The results are averaged over three datasets.}
    \label{fig:ablation-bi}
\end{figure}

We further conduct ablation studies to validate the effectiveness of the bi-level optimization framework. Specifically, we compare \model{} with Separate Training (Separate). Separate uses the same fairness-aware recommendation loss as \model{}. It first trains the recommendation model independently and then freezes it before training the representation. As shown in Figure ~\ref{fig:ablation-bi}, \model{} consistently outperforms Separate in both recommendation performance and fairness metrics. The parameters in Separate are optimized in isolation, lacking mutual influence during training, which limits its ability to simultaneously mitigate prior unfairness and training-time unfairness. In contrast, \model{} simultaneously optimizes both components, allowing fairness improvements in both dimensions to reinforce each other, ultimately enhancing the overall recommendation performance.
\begin{table}
\centering
\caption{Comparison of \model{} with the best-performing baseline under different base LLMs on Games Dataset.}
\label{tab:llm}
\resizebox{\linewidth}{!}{
\begin{tabular}{cl|ccc|cc|cc}
    \toprule
    \multirow{2}{*}{base LLM} & \multicolumn{1}{c}{\multirow{2}{*}{Method}} & \multicolumn{3}{|c|}{Accuracy} & \multicolumn{2}{c|}{Pop. Fairness} & \multicolumn{2}{c}{Genre Fairness} \\
          &       & Recall$\uparrow$ & NDCG$\uparrow$  & HR$\uparrow$    & CV$\downarrow$ & MIN$\uparrow$ & CV$\downarrow$ & MIN$\uparrow$ \\
    \midrule
    \multirow{3}[1]{*}{Qwen} & AlphaRec & 0.1324 & 0.1021 & 0.2939 & 0.5210 & 0.0578 & 0.5454 & 0.0267 \\
          & DRORec & \textbf{0.1367} & \textbf{0.1033} & \textbf{0.2989} & 0.5094 & 0.0593 & 0.5236 & 0.0271 \\
          & \model{} & 0.1346 & 0.1015 & 0.2951 & \textbf{0.5053} & \textbf{0.0606} & \textbf{0.5189} & \textbf{0.0279} \\
    \midrule
    \multirow{3}[1]{*}{Llama} & AlphaRec & \textbf{0.1601} & \textbf{0.1159} & \textbf{0.3406} & 0.4668 & 0.0782 & 0.4152 & 0.0630 \\
          & DRORec & 0.1524 & 0.1112 & 0.3268 & 0.4254 & 0.0810 & 0.4215 & 0.0607 \\
          & \model{} & 0.1564 & 0.1134 & 0.3326 & \textbf{0.3997} & \textbf{0.0863} & \textbf{0.3983} & \textbf{0.0688} \\
    \bottomrule
    \end{tabular}%
    }
\end{table}

\begin{table}
\centering
\caption{Comparison of \model{} with the best-performing baseline for Linear on Movies Dataset.}
\label{tab:lrs}
\resizebox{\linewidth}{!}{
\begin{tabular}{l|ccc|cc|cc}
    \toprule
    \multicolumn{1}{c}{\multirow{2}{*}{Method}} & \multicolumn{3}{|c|}{Accuracy} & \multicolumn{2}{c|}{Pop. Fairness} & \multicolumn{2}{c}{Genre Fairness} \\
          & \multicolumn{1}{c}{Recall$\uparrow$} & \multicolumn{1}{c}{NDCG$\uparrow$} & \multicolumn{1}{c|}{HR$\uparrow$} & \multicolumn{1}{c}{CV$\downarrow$} & \multicolumn{1}{c|}{MIN$\uparrow$} & \multicolumn{1}{c}{CV$\downarrow$} & \multicolumn{1}{c}{MIN$\uparrow$} \\
    \midrule
    Linear & 0.1969 & 0.1750 & 0.4194 & 0.9643 & 0.0106 & 0.2673 & 0.1224 \\
    ITFR  & 0.1975 & 0.1759 & 0.4183 & 0.9538 & 0.0135 & 0.2400 & 0.1289 \\
    \model{} & \textbf{0.1986} & \textbf{0.1776} & \textbf{0.4235} & \textbf{0.9407} & \textbf{0.0171} & \textbf{0.2245} & \textbf{0.1310} \\
    \bottomrule
    \end{tabular}%
    }
\end{table}
\section{RELATED WORK}
\subsection{LLMs for Recommender Systems}
The application of LLMs in RSs can be broadly categorized into LLM-based RSs and LLM-enhanced RSs.
LLM-based RSs replace conventional recommendation models with LLMs as a novel backbone. 
In this paradigm, effective prompt templates are designed~\cite{lyu2023llm}, and fine-tuning techniques commonly used for LLMs, such as supervised fine-tuning~\cite{bao2023bi} or preference optimization~\cite{bai2024aligning,liao2024rosepo}, are applied to align the LLM’s generative objective with recommendation tasks. 
This paradigm is predominantly used in sequential recommendation scenarios.
LLM-enhanced RSs leverage LLMs, which have been pretrained on massive text corpora, to generate powerful and expressive textual representations that assist recommendation models. 
The LLM input can include item titles, item descriptions, user intents, and other metadata~\cite{ren2024representation,wei2024llmrec}. 
Since LLM-generated representations inherently encode user preference signals~\cite{sheng2024language}, integrating them into recommendation models significantly boosts recommendation performance.

While the introduction of LLMs has led to notable improvements in recommendation performance, the inherent biases within LLMs also introduce additional fairness concerns into RSs.

\subsection{Item-side Fairness in Recommender Systems}
Fairness in RSs is typically categorized into individual fairness~\cite{dwork2012fairness,wang2022providing} and group fairness~\cite{stratigi2020fair,wang2022make}, with the latter often defined based on attributes such as item popularity or item category. 
Since LLMs primarily leverage item metadata, our study focuses on group fairness from the item side, advocating for equitable treatment across different item groups.
In traditional RSs, fairness is often addressed through a three-stage approach: i) adjusting data distributions before training~\cite{ekstrand2018all,rastegarpanah2019fighting}, ii) modifying the loss function or introducing regularization terms during training~\cite{burke2018balanced,beutel2019fairness,wu2022multi}, and iii) adjusting the ranking process after training~\cite{patro2022fair,zehlike2022fairness}.
Among these, pre-training and post-training approaches can be directly incorporated into LLM-enhanced RSs.

Notably, in the context of LLM-based RSs, existing studies have highlighted fairness issues and proposed corresponding solutions. 
For instance, \cite{jiang2024item} reweights samples during fine-tuning or re-ranks recommendation results to mitigate unfairness.
Additionally, DPO-based optimization has been explored as a means to achieve better fairness alignment~\cite{gao2024sprec}.

However, current research on IF in LLM-based recommendation and its mitigation strategies primarily focus on sequential recommendation, and no general solution has been proposed for LLM-enhanced general RSs. 
Furthermore, existing methods overlook the hard-to-estimate prior unfairness introduced by LLMs, which makes it challenging to directly apply in-training fairness approaches from traditional RSs.
\section{CONCLUSION}
In this paper, we conduct the first systematic study on fairness issues in LLM-enhanced RSs.  
Surprisingly, our empirical findings reveal that LLM-enhanced RSs can achieve better fairness than traditional methods.
To improve fairness in LLM-enhanced RSs, we identify two major sources of unfairness: i) prior unfairness from LLM-generated item representations and ii) training unfairness from the downstream recommendation model. 
Thus, we propose \model{}, a fairness-aware training framework based on bi-level optimization. 
\model{} simultaneously mitigates prior and training unfairness by optimizing two sets of parameters: LLM-generated item representations and the recommendation model's projection layer.  
Additionally, we introduce an adaptive inter-group balancing mechanism, which aligns gradient directions benefiting all groups and dynamically balances their performance during training.
Extensive experiments on multiple real-world datasets validate the effectiveness of \model{}, demonstrating significant improvements in fairness.
%
Our work sheds light on the fairness implications of integrating LLMs into RSs and offers actionable insights for addressing these challenges. 

\bibliographystyle{ACM-Reference-Format}
\bibliography{author_draft}


\begin{thebibliography}{62}


\ifx \showCODEN    \undefined \def \showCODEN     #1{\unskip}     \fi
\ifx \showISBNx    \undefined \def \showISBNx     #1{\unskip}     \fi
\ifx \showISBNxiii \undefined \def \showISBNxiii  #1{\unskip}     \fi
\ifx \showISSN     \undefined \def \showISSN      #1{\unskip}     \fi
\ifx \showLCCN     \undefined \def \showLCCN      #1{\unskip}     \fi
\ifx \shownote     \undefined \def \shownote      #1{#1}          \fi
\ifx \showarticletitle \undefined \def \showarticletitle #1{#1}   \fi
\ifx \showURL      \undefined \def \showURL       {\relax}        \fi
\providecommand\bibfield[2]{#2}
\providecommand\bibinfo[2]{#2}
\providecommand\natexlab[1]{#1}
\providecommand\showeprint[2][]{arXiv:#2}

\bibitem[Achiam et~al\mbox{.}(2023)]%
        {achiam2023gpt}
\bibfield{author}{\bibinfo{person}{Josh Achiam}, \bibinfo{person}{Steven Adler}, \bibinfo{person}{Sandhini Agarwal}, \bibinfo{person}{Lama Ahmad}, \bibinfo{person}{Ilge Akkaya}, \bibinfo{person}{Florencia~Leoni Aleman}, \bibinfo{person}{Diogo Almeida}, \bibinfo{person}{Janko Altenschmidt}, \bibinfo{person}{Sam Altman}, \bibinfo{person}{Shyamal Anadkat}, {et~al\mbox{.}}} \bibinfo{year}{2023}\natexlab{}.
\newblock \showarticletitle{Gpt-4 technical report}.
\newblock \bibinfo{journal}{\emph{arXiv preprint arXiv:2303.08774}} (\bibinfo{year}{2023}).
\newblock


\bibitem[Bai et~al\mbox{.}(2024a)]%
        {bai2024measuring}
\bibfield{author}{\bibinfo{person}{Xuechunzi Bai}, \bibinfo{person}{Angelina Wang}, \bibinfo{person}{Ilia Sucholutsky}, {and} \bibinfo{person}{Thomas~L Griffiths}.} \bibinfo{year}{2024}\natexlab{a}.
\newblock \showarticletitle{Measuring Implicit Bias in Explicitly Unbiased Large Language Models}. In \bibinfo{booktitle}{\emph{NeurIPS 2024 Workshop on Behavioral Machine Learning}}.
\newblock


\bibitem[Bai et~al\mbox{.}(2025)]%
        {bai2025explicitly}
\bibfield{author}{\bibinfo{person}{Xuechunzi Bai}, \bibinfo{person}{Angelina Wang}, \bibinfo{person}{Ilia Sucholutsky}, {and} \bibinfo{person}{Thomas~L Griffiths}.} \bibinfo{year}{2025}\natexlab{}.
\newblock \showarticletitle{Explicitly unbiased large language models still form biased associations}.
\newblock \bibinfo{journal}{\emph{Proceedings of the National Academy of Sciences}} \bibinfo{volume}{122}, \bibinfo{number}{8} (\bibinfo{year}{2025}), \bibinfo{pages}{e2416228122}.
\newblock


\bibitem[Bai et~al\mbox{.}(2024b)]%
        {bai2024aligning}
\bibfield{author}{\bibinfo{person}{Zhuoxi Bai}, \bibinfo{person}{Ning Wu}, \bibinfo{person}{Fengyu Cai}, \bibinfo{person}{Xinyi Zhu}, {and} \bibinfo{person}{Yun Xiong}.} \bibinfo{year}{2024}\natexlab{b}.
\newblock \showarticletitle{Aligning Large Language Model with Direct Multi-Preference Optimization for Recommendation}. In \bibinfo{booktitle}{\emph{Proceedings of the 33rd CIKM}}. \bibinfo{pages}{76--86}.
\newblock


\bibitem[Bao et~al\mbox{.}(2023)]%
        {bao2023bi}
\bibfield{author}{\bibinfo{person}{Keqin Bao}, \bibinfo{person}{Jizhi Zhang}, \bibinfo{person}{Wenjie Wang}, \bibinfo{person}{Yang Zhang}, \bibinfo{person}{Zhengyi Yang}, \bibinfo{person}{Yancheng Luo}, \bibinfo{person}{Chong Chen}, \bibinfo{person}{Fuli Feng}, {and} \bibinfo{person}{Qi Tian}.} \bibinfo{year}{2023}\natexlab{}.
\newblock \showarticletitle{A bi-step grounding paradigm for large language models in recommendation systems}.
\newblock \bibinfo{journal}{\emph{arXiv preprint arXiv:2308.08434}} (\bibinfo{year}{2023}).
\newblock


\bibitem[Bao and Jiang(2016)]%
        {bao2016intelligent}
\bibfield{author}{\bibinfo{person}{Youjun Bao} {and} \bibinfo{person}{Xiaohong Jiang}.} \bibinfo{year}{2016}\natexlab{}.
\newblock \showarticletitle{An intelligent medicine recommender system framework}. In \bibinfo{booktitle}{\emph{2016 IEEE 11Th conference on industrial electronics and applications (ICIEA)}}. IEEE, \bibinfo{pages}{1383--1388}.
\newblock


\bibitem[Beutel et~al\mbox{.}(2019)]%
        {beutel2019fairness}
\bibfield{author}{\bibinfo{person}{Alex Beutel}, \bibinfo{person}{Jilin Chen}, \bibinfo{person}{Tulsee Doshi}, \bibinfo{person}{Hai Qian}, \bibinfo{person}{Li Wei}, \bibinfo{person}{Yi Wu}, \bibinfo{person}{Lukasz Heldt}, \bibinfo{person}{Zhe Zhao}, \bibinfo{person}{Lichan Hong}, \bibinfo{person}{Ed~H Chi}, {et~al\mbox{.}}} \bibinfo{year}{2019}\natexlab{}.
\newblock \showarticletitle{Fairness in recommendation ranking through pairwise comparisons}. In \bibinfo{booktitle}{\emph{Proceedings of the 25th ACM SIGKDD international conference on knowledge discovery \& data mining}}. \bibinfo{pages}{2212--2220}.
\newblock


\bibitem[Burke et~al\mbox{.}(2018)]%
        {burke2018balanced}
\bibfield{author}{\bibinfo{person}{Robin Burke}, \bibinfo{person}{Nasim Sonboli}, {and} \bibinfo{person}{Aldo Ordonez-Gauger}.} \bibinfo{year}{2018}\natexlab{}.
\newblock \showarticletitle{Balanced neighborhoods for multi-sided fairness in recommendation}. In \bibinfo{booktitle}{\emph{Conference on fairness, accountability and transparency}}. PMLR, \bibinfo{pages}{202--214}.
\newblock


\bibitem[Chen et~al\mbox{.}(2023)]%
        {chen2023fairly}
\bibfield{author}{\bibinfo{person}{Xiao Chen}, \bibinfo{person}{Wenqi Fan}, \bibinfo{person}{Jingfan Chen}, \bibinfo{person}{Haochen Liu}, \bibinfo{person}{Zitao Liu}, \bibinfo{person}{Zhaoxiang Zhang}, {and} \bibinfo{person}{Qing Li}.} \bibinfo{year}{2023}\natexlab{}.
\newblock \showarticletitle{Fairly adaptive negative sampling for recommendations}. In \bibinfo{booktitle}{\emph{Proceedings of the ACM Web Conference 2023}}. \bibinfo{pages}{3723--3733}.
\newblock


\bibitem[D{\'e}sid{\'e}ri(2012)]%
        {desideri2012multiple}
\bibfield{author}{\bibinfo{person}{Jean-Antoine D{\'e}sid{\'e}ri}.} \bibinfo{year}{2012}\natexlab{}.
\newblock \showarticletitle{Multiple-gradient descent algorithm (MGDA) for multiobjective optimization}.
\newblock \bibinfo{journal}{\emph{Comptes Rendus Mathematique}} \bibinfo{volume}{350}, \bibinfo{number}{5-6} (\bibinfo{year}{2012}), \bibinfo{pages}{313--318}.
\newblock


\bibitem[Dwork et~al\mbox{.}(2012)]%
        {dwork2012fairness}
\bibfield{author}{\bibinfo{person}{Cynthia Dwork}, \bibinfo{person}{Moritz Hardt}, \bibinfo{person}{Toniann Pitassi}, \bibinfo{person}{Omer Reingold}, {and} \bibinfo{person}{Richard Zemel}.} \bibinfo{year}{2012}\natexlab{}.
\newblock \showarticletitle{Fairness through awareness}. In \bibinfo{booktitle}{\emph{Proceedings of the 3rd innovations in theoretical computer science conference}}. \bibinfo{pages}{214--226}.
\newblock


\bibitem[Ekstrand et~al\mbox{.}(2018)]%
        {ekstrand2018all}
\bibfield{author}{\bibinfo{person}{Michael~D Ekstrand}, \bibinfo{person}{Mucun Tian}, \bibinfo{person}{Ion~Madrazo Azpiazu}, \bibinfo{person}{Jennifer~D Ekstrand}, \bibinfo{person}{Oghenemaro Anuyah}, \bibinfo{person}{David McNeill}, {and} \bibinfo{person}{Maria~Soledad Pera}.} \bibinfo{year}{2018}\natexlab{}.
\newblock \showarticletitle{All the cool kids, how do they fit in?: Popularity and demographic biases in recommender evaluation and effectiveness}. In \bibinfo{booktitle}{\emph{Conference on fairness, accountability and transparency}}. PMLR, \bibinfo{pages}{172--186}.
\newblock


\bibitem[Gao et~al\mbox{.}(2024)]%
        {gao2024sprec}
\bibfield{author}{\bibinfo{person}{Chongming Gao}, \bibinfo{person}{Ruijun Chen}, \bibinfo{person}{Shuai Yuan}, \bibinfo{person}{Kexin Huang}, \bibinfo{person}{Yuanqing Yu}, {and} \bibinfo{person}{Xiangnan He}.} \bibinfo{year}{2024}\natexlab{}.
\newblock \showarticletitle{SPRec: Leveraging Self-Play to Debias Preference Alignment for Large Language Model-based Recommendations}.
\newblock \bibinfo{journal}{\emph{arXiv preprint arXiv:2412.09243}} (\bibinfo{year}{2024}).
\newblock


\bibitem[Han et~al\mbox{.}(2023)]%
        {han2023processing}
\bibfield{author}{\bibinfo{person}{Zhongxuan Han}, \bibinfo{person}{Chaochao Chen}, \bibinfo{person}{Xiaolin Zheng}, \bibinfo{person}{Weiming Liu}, \bibinfo{person}{Jun Wang}, \bibinfo{person}{Wenjie Cheng}, {and} \bibinfo{person}{Yuyuan Li}.} \bibinfo{year}{2023}\natexlab{}.
\newblock \showarticletitle{In-processing user constrained dominant sets for user-oriented fairness in recommender systems}. In \bibinfo{booktitle}{\emph{Proceedings of the 31st ACM International Conference on Multimedia}}. \bibinfo{pages}{6190--6201}.
\newblock


\bibitem[He et~al\mbox{.}(2020)]%
        {he2020lightgcn}
\bibfield{author}{\bibinfo{person}{Xiangnan He}, \bibinfo{person}{Kuan Deng}, \bibinfo{person}{Xiang Wang}, \bibinfo{person}{Yan Li}, \bibinfo{person}{Yongdong Zhang}, {and} \bibinfo{person}{Meng Wang}.} \bibinfo{year}{2020}\natexlab{}.
\newblock \showarticletitle{Lightgcn: Simplifying and powering graph convolution network for recommendation}. In \bibinfo{booktitle}{\emph{Proceedings of the 43rd International ACM SIGIR conference on research and development in Information Retrieval}}. \bibinfo{pages}{639--648}.
\newblock


\bibitem[Hu et~al\mbox{.}(2008)]%
        {hu2008collaborative}
\bibfield{author}{\bibinfo{person}{Yifan Hu}, \bibinfo{person}{Yehuda Koren}, {and} \bibinfo{person}{Chris Volinsky}.} \bibinfo{year}{2008}\natexlab{}.
\newblock \showarticletitle{Collaborative filtering for implicit feedback datasets}. In \bibinfo{booktitle}{\emph{2008 Eighth IEEE international conference on data mining}}. Ieee, \bibinfo{pages}{263--272}.
\newblock


\bibitem[Jaggi(2013)]%
        {jaggi2013revisiting}
\bibfield{author}{\bibinfo{person}{Martin Jaggi}.} \bibinfo{year}{2013}\natexlab{}.
\newblock \showarticletitle{Revisiting Frank-Wolfe: Projection-free sparse convex optimization}. In \bibinfo{booktitle}{\emph{International conference on machine learning}}. PMLR, \bibinfo{pages}{427--435}.
\newblock


\bibitem[Jiang et~al\mbox{.}(2024)]%
        {jiang2024item}
\bibfield{author}{\bibinfo{person}{Meng Jiang}, \bibinfo{person}{Keqin Bao}, \bibinfo{person}{Jizhi Zhang}, \bibinfo{person}{Wenjie Wang}, \bibinfo{person}{Zhengyi Yang}, \bibinfo{person}{Fuli Feng}, {and} \bibinfo{person}{Xiangnan He}.} \bibinfo{year}{2024}\natexlab{}.
\newblock \showarticletitle{Item-side Fairness of Large Language Model-based Recommendation System}. In \bibinfo{booktitle}{\emph{Proceedings of the ACM on Web Conference 2024}}. \bibinfo{pages}{4717--4726}.
\newblock


\bibitem[Koren et~al\mbox{.}(2009)]%
        {koren2009matrix}
\bibfield{author}{\bibinfo{person}{Yehuda Koren}, \bibinfo{person}{Robert Bell}, {and} \bibinfo{person}{Chris Volinsky}.} \bibinfo{year}{2009}\natexlab{}.
\newblock \showarticletitle{Matrix factorization techniques for recommender systems}.
\newblock \bibinfo{journal}{\emph{Computer}} \bibinfo{volume}{42}, \bibinfo{number}{8} (\bibinfo{year}{2009}), \bibinfo{pages}{30--37}.
\newblock


\bibitem[Krichene and Rendle(2020)]%
        {krichene2020sampled}
\bibfield{author}{\bibinfo{person}{Walid Krichene} {and} \bibinfo{person}{Steffen Rendle}.} \bibinfo{year}{2020}\natexlab{}.
\newblock \showarticletitle{On sampled metrics for item recommendation}. In \bibinfo{booktitle}{\emph{Proceedings of the 26th ACM SIGKDD international conference on knowledge discovery \& data mining}}. \bibinfo{pages}{1748--1757}.
\newblock


\bibitem[Li et~al\mbox{.}(2021b)]%
        {li2021leave}
\bibfield{author}{\bibinfo{person}{Roger~Zhe Li}, \bibinfo{person}{Juli{\'a}n Urbano}, {and} \bibinfo{person}{Alan Hanjalic}.} \bibinfo{year}{2021}\natexlab{b}.
\newblock \showarticletitle{Leave no user behind: Towards improving the utility of recommender systems for non-mainstream users}. In \bibinfo{booktitle}{\emph{Proceedings of the 14th ACM International Conference on Web Search and Data Mining}}. \bibinfo{pages}{103--111}.
\newblock


\bibitem[Li et~al\mbox{.}(2021a)]%
        {li2021user}
\bibfield{author}{\bibinfo{person}{Yunqi Li}, \bibinfo{person}{Hanxiong Chen}, \bibinfo{person}{Zuohui Fu}, \bibinfo{person}{Yingqiang Ge}, {and} \bibinfo{person}{Yongfeng Zhang}.} \bibinfo{year}{2021}\natexlab{a}.
\newblock \showarticletitle{User-oriented fairness in recommendation}. In \bibinfo{booktitle}{\emph{Proceedings of the web conference 2021}}. \bibinfo{pages}{624--632}.
\newblock


\bibitem[Li et~al\mbox{.}(2023)]%
        {li2023fairness}
\bibfield{author}{\bibinfo{person}{Yunqi Li}, \bibinfo{person}{Hanxiong Chen}, \bibinfo{person}{Shuyuan Xu}, \bibinfo{person}{Yingqiang Ge}, \bibinfo{person}{Juntao Tan}, \bibinfo{person}{Shuchang Liu}, {and} \bibinfo{person}{Yongfeng Zhang}.} \bibinfo{year}{2023}\natexlab{}.
\newblock \showarticletitle{Fairness in recommendation: Foundations, methods, and applications}.
\newblock \bibinfo{journal}{\emph{ACM Transactions on Intelligent Systems and Technology}} \bibinfo{volume}{14}, \bibinfo{number}{5} (\bibinfo{year}{2023}), \bibinfo{pages}{1--48}.
\newblock


\bibitem[Liang et~al\mbox{.}(2018)]%
        {liang2018variational}
\bibfield{author}{\bibinfo{person}{Dawen Liang}, \bibinfo{person}{Rahul~G Krishnan}, \bibinfo{person}{Matthew~D Hoffman}, {and} \bibinfo{person}{Tony Jebara}.} \bibinfo{year}{2018}\natexlab{}.
\newblock \showarticletitle{Variational autoencoders for collaborative filtering}. In \bibinfo{booktitle}{\emph{Proceedings of the 2018 world wide web conference}}. \bibinfo{pages}{689--698}.
\newblock


\bibitem[Liao et~al\mbox{.}(2024a)]%
        {liao2024rosepo}
\bibfield{author}{\bibinfo{person}{Jiayi Liao}, \bibinfo{person}{Xiangnan He}, \bibinfo{person}{Ruobing Xie}, \bibinfo{person}{Jiancan Wu}, \bibinfo{person}{Yancheng Yuan}, \bibinfo{person}{Xingwu Sun}, \bibinfo{person}{Zhanhui Kang}, {and} \bibinfo{person}{Xiang Wang}.} \bibinfo{year}{2024}\natexlab{a}.
\newblock \showarticletitle{RosePO: Aligning LLM-based Recommenders with Human Values}.
\newblock \bibinfo{journal}{\emph{arXiv preprint arXiv:2410.12519}} (\bibinfo{year}{2024}).
\newblock


\bibitem[Liao et~al\mbox{.}(2024b)]%
        {liao2024llara}
\bibfield{author}{\bibinfo{person}{Jiayi Liao}, \bibinfo{person}{Sihang Li}, \bibinfo{person}{Zhengyi Yang}, \bibinfo{person}{Jiancan Wu}, \bibinfo{person}{Yancheng Yuan}, \bibinfo{person}{Xiang Wang}, {and} \bibinfo{person}{Xiangnan He}.} \bibinfo{year}{2024}\natexlab{b}.
\newblock \showarticletitle{Llara: Large language-recommendation assistant}. In \bibinfo{booktitle}{\emph{Proceedings of the 47th International ACM SIGIR Conference on Research and Development in Information Retrieval}}. \bibinfo{pages}{1785--1795}.
\newblock


\bibitem[Liu et~al\mbox{.}(2019)]%
        {liudarts}
\bibfield{author}{\bibinfo{person}{Hanxiao Liu}, \bibinfo{person}{Karen Simonyan}, {and} \bibinfo{person}{Yiming Yang}.} \bibinfo{year}{2019}\natexlab{}.
\newblock \showarticletitle{DARTS: Differentiable Architecture Search}. In \bibinfo{booktitle}{\emph{International Conference on Learning Representations}}.
\newblock


\bibitem[Liu et~al\mbox{.}(2021)]%
        {liutowards}
\bibfield{author}{\bibinfo{person}{Liyang Liu}, \bibinfo{person}{Yi Li}, \bibinfo{person}{Zhanghui Kuang}, \bibinfo{person}{Jing-Hao Xue}, \bibinfo{person}{Yimin Chen}, \bibinfo{person}{Wenming Yang}, \bibinfo{person}{Qingmin Liao}, {and} \bibinfo{person}{Wayne Zhang}.} \bibinfo{year}{2021}\natexlab{}.
\newblock \showarticletitle{Towards Impartial Multi-task Learning}. In \bibinfo{booktitle}{\emph{International Conference on Learning Representations}}.
\newblock


\bibitem[Lyu et~al\mbox{.}(2023)]%
        {lyu2023llm}
\bibfield{author}{\bibinfo{person}{Hanjia Lyu}, \bibinfo{person}{Song Jiang}, \bibinfo{person}{Hanqing Zeng}, \bibinfo{person}{Yinglong Xia}, \bibinfo{person}{Qifan Wang}, \bibinfo{person}{Si Zhang}, \bibinfo{person}{Ren Chen}, \bibinfo{person}{Christopher Leung}, \bibinfo{person}{Jiajie Tang}, {and} \bibinfo{person}{Jiebo Luo}.} \bibinfo{year}{2023}\natexlab{}.
\newblock \showarticletitle{Llm-rec: Personalized recommendation via prompting large language models}.
\newblock \bibinfo{journal}{\emph{arXiv preprint arXiv:2307.15780}} (\bibinfo{year}{2023}).
\newblock


\bibitem[Meng et~al\mbox{.}(2024)]%
        {meng2024sfrembedding}
\bibfield{author}{\bibinfo{person}{Rui Meng}, \bibinfo{person}{Ye Liu}, \bibinfo{person}{Shafiq~Rayhan Joty}, \bibinfo{person}{Caiming Xiong}, \bibinfo{person}{Yingbo Zhou}, {and} \bibinfo{person}{Semih Yavuz}.} \bibinfo{year}{2024}\natexlab{}.
\newblock \showarticletitle{Sfrembedding-mistral: enhance text retrieval with transfer learning}.
\newblock \bibinfo{journal}{\emph{Salesforce AI Research Blog}}  \bibinfo{volume}{3} (\bibinfo{year}{2024}), \bibinfo{pages}{6}.
\newblock


\bibitem[Oord et~al\mbox{.}(2018)]%
        {oord2018representation}
\bibfield{author}{\bibinfo{person}{Aaron van~den Oord}, \bibinfo{person}{Yazhe Li}, {and} \bibinfo{person}{Oriol Vinyals}.} \bibinfo{year}{2018}\natexlab{}.
\newblock \showarticletitle{Representation learning with contrastive predictive coding}.
\newblock \bibinfo{journal}{\emph{arXiv preprint arXiv:1807.03748}} (\bibinfo{year}{2018}).
\newblock


\bibitem[Paparrizos et~al\mbox{.}(2011)]%
        {paparrizos2011machine}
\bibfield{author}{\bibinfo{person}{Ioannis Paparrizos}, \bibinfo{person}{B~Barla Cambazoglu}, {and} \bibinfo{person}{Aristides Gionis}.} \bibinfo{year}{2011}\natexlab{}.
\newblock \showarticletitle{Machine learned job recommendation}. In \bibinfo{booktitle}{\emph{Proceedings of the fifth ACM Conference on Recommender Systems}}. \bibinfo{pages}{325--328}.
\newblock


\bibitem[Patro et~al\mbox{.}(2022)]%
        {patro2022fair}
\bibfield{author}{\bibinfo{person}{Gourab~K Patro}, \bibinfo{person}{Lorenzo Porcaro}, \bibinfo{person}{Laura Mitchell}, \bibinfo{person}{Qiuyue Zhang}, \bibinfo{person}{Meike Zehlike}, {and} \bibinfo{person}{Nikhil Garg}.} \bibinfo{year}{2022}\natexlab{}.
\newblock \showarticletitle{Fair ranking: a critical review, challenges, and future directions}. In \bibinfo{booktitle}{\emph{Proceedings of the 2022 ACM conference on fairness, accountability, and transparency}}. \bibinfo{pages}{1929--1942}.
\newblock


\bibitem[Rastegarpanah et~al\mbox{.}(2019)]%
        {rastegarpanah2019fighting}
\bibfield{author}{\bibinfo{person}{Bashir Rastegarpanah}, \bibinfo{person}{Krishna~P Gummadi}, {and} \bibinfo{person}{Mark Crovella}.} \bibinfo{year}{2019}\natexlab{}.
\newblock \showarticletitle{Fighting fire with fire: Using antidote data to improve polarization and fairness of recommender systems}. In \bibinfo{booktitle}{\emph{Proceedings of the twelfth ACM international conference on web search and data mining}}. \bibinfo{pages}{231--239}.
\newblock


\bibitem[Ren et~al\mbox{.}(2024)]%
        {ren2024representation}
\bibfield{author}{\bibinfo{person}{Xubin Ren}, \bibinfo{person}{Wei Wei}, \bibinfo{person}{Lianghao Xia}, \bibinfo{person}{Lixin Su}, \bibinfo{person}{Suqi Cheng}, \bibinfo{person}{Junfeng Wang}, \bibinfo{person}{Dawei Yin}, {and} \bibinfo{person}{Chao Huang}.} \bibinfo{year}{2024}\natexlab{}.
\newblock \showarticletitle{Representation learning with large language models for recommendation}. In \bibinfo{booktitle}{\emph{Proceedings of the ACM Web Conference 2024}}. \bibinfo{pages}{3464--3475}.
\newblock


\bibitem[Rendle et~al\mbox{.}(2009)]%
        {rendle2009bpr}
\bibfield{author}{\bibinfo{person}{Steffen Rendle}, \bibinfo{person}{Christoph Freudenthaler}, \bibinfo{person}{Zeno Gantner}, {and} \bibinfo{person}{Lars Schmidt-Thieme}.} \bibinfo{year}{2009}\natexlab{}.
\newblock \showarticletitle{BPR: Bayesian personalized ranking from implicit feedback}. In \bibinfo{booktitle}{\emph{Proceedings of the Twenty-Fifth Conference on Uncertainty in Artificial Intelligence}}. \bibinfo{pages}{452--461}.
\newblock


\bibitem[Sagawa et~al\mbox{.}(2019)]%
        {sagawadistributionally}
\bibfield{author}{\bibinfo{person}{Shiori Sagawa}, \bibinfo{person}{Pang~Wei Koh}, \bibinfo{person}{Tatsunori~B Hashimoto}, {and} \bibinfo{person}{Percy Liang}.} \bibinfo{year}{2019}\natexlab{}.
\newblock \showarticletitle{Distributionally Robust Neural Networks}. In \bibinfo{booktitle}{\emph{International Conference on Learning Representations}}.
\newblock


\bibitem[Sheng et~al\mbox{.}(2024)]%
        {sheng2024language}
\bibfield{author}{\bibinfo{person}{Leheng Sheng}, \bibinfo{person}{An Zhang}, \bibinfo{person}{Yi Zhang}, \bibinfo{person}{Yuxin Chen}, \bibinfo{person}{Xiang Wang}, {and} \bibinfo{person}{Tat-Seng Chua}.} \bibinfo{year}{2024}\natexlab{}.
\newblock \showarticletitle{Language Representations Can be What Recommenders Need: Findings and Potentials}.
\newblock \bibinfo{journal}{\emph{arXiv preprint arXiv:2407.05441}} (\bibinfo{year}{2024}).
\newblock


\bibitem[Stratigi et~al\mbox{.}(2020)]%
        {stratigi2020fair}
\bibfield{author}{\bibinfo{person}{Maria Stratigi}, \bibinfo{person}{Jyrki Nummenmaa}, \bibinfo{person}{Evaggelia Pitoura}, {and} \bibinfo{person}{Kostas Stefanidis}.} \bibinfo{year}{2020}\natexlab{}.
\newblock \showarticletitle{Fair sequential group recommendations}. In \bibinfo{booktitle}{\emph{Proceedings of the 35th Annual ACM Symposium on Applied Computing}}. \bibinfo{pages}{1443--1452}.
\newblock


\bibitem[Tommasel(2024)]%
        {tommasel2024fairness}
\bibfield{author}{\bibinfo{person}{Antonela Tommasel}.} \bibinfo{year}{2024}\natexlab{}.
\newblock \showarticletitle{Fairness Matters: A look at LLM-generated group recommendations}. In \bibinfo{booktitle}{\emph{Proceedings of the 18th ACM Conference on Recommender Systems}}. \bibinfo{pages}{993--998}.
\newblock


\bibitem[Touvron et~al\mbox{.}(2023)]%
        {touvron2023llama}
\bibfield{author}{\bibinfo{person}{Hugo Touvron}, \bibinfo{person}{Thibaut Lavril}, \bibinfo{person}{Gautier Izacard}, \bibinfo{person}{Xavier Martinet}, \bibinfo{person}{Marie-Anne Lachaux}, \bibinfo{person}{Timoth{\'e}e Lacroix}, \bibinfo{person}{Baptiste Rozi{\`e}re}, \bibinfo{person}{Naman Goyal}, \bibinfo{person}{Eric Hambro}, \bibinfo{person}{Faisal Azhar}, {et~al\mbox{.}}} \bibinfo{year}{2023}\natexlab{}.
\newblock \showarticletitle{Llama: Open and efficient foundation language models}.
\newblock \bibinfo{journal}{\emph{arXiv preprint arXiv:2302.13971}} (\bibinfo{year}{2023}).
\newblock


\bibitem[Turcotte et~al\mbox{.}(2015)]%
        {turcotte2015news}
\bibfield{author}{\bibinfo{person}{Jason Turcotte}, \bibinfo{person}{Chance York}, \bibinfo{person}{Jacob Irving}, \bibinfo{person}{Rosanne~M Scholl}, {and} \bibinfo{person}{Raymond~J Pingree}.} \bibinfo{year}{2015}\natexlab{}.
\newblock \showarticletitle{News recommendations from social media opinion leaders: Effects on media trust and information seeking}.
\newblock \bibinfo{journal}{\emph{Journal of computer-mediated communication}} \bibinfo{volume}{20}, \bibinfo{number}{5} (\bibinfo{year}{2015}), \bibinfo{pages}{520--535}.
\newblock


\bibitem[Wang et~al\mbox{.}(2022)]%
        {wang2022make}
\bibfield{author}{\bibinfo{person}{Jiayin Wang}, \bibinfo{person}{Weizhi Ma}, \bibinfo{person}{Jiayu Li}, \bibinfo{person}{Hongyu Lu}, \bibinfo{person}{Min Zhang}, \bibinfo{person}{Biao Li}, \bibinfo{person}{Yiqun Liu}, \bibinfo{person}{Peng Jiang}, {and} \bibinfo{person}{Shaoping Ma}.} \bibinfo{year}{2022}\natexlab{}.
\newblock \showarticletitle{Make fairness more fair: Fair item utility estimation and exposure re-distribution}. In \bibinfo{booktitle}{\emph{Proceedings of the 28th ACM SIGKDD Conference on Knowledge Discovery and Data Mining}}. \bibinfo{pages}{1868--1877}.
\newblock


\bibitem[Wang and Wang(2022)]%
        {wang2022providing}
\bibfield{author}{\bibinfo{person}{Xiuling Wang} {and} \bibinfo{person}{Wendy~Hui Wang}.} \bibinfo{year}{2022}\natexlab{}.
\newblock \showarticletitle{Providing item-side individual fairness for deep recommender systems}. In \bibinfo{booktitle}{\emph{Proceedings of the 2022 ACM Conference on Fairness, Accountability, and Transparency}}. \bibinfo{pages}{117--127}.
\newblock


\bibitem[Wang et~al\mbox{.}(2023)]%
        {wang2023survey}
\bibfield{author}{\bibinfo{person}{Yifan Wang}, \bibinfo{person}{Weizhi Ma}, \bibinfo{person}{Min Zhang}, \bibinfo{person}{Yiqun Liu}, {and} \bibinfo{person}{Shaoping Ma}.} \bibinfo{year}{2023}\natexlab{}.
\newblock \showarticletitle{A survey on the fairness of recommender systems}.
\newblock \bibinfo{journal}{\emph{ACM Transactions on Information Systems}} \bibinfo{volume}{41}, \bibinfo{number}{3} (\bibinfo{year}{2023}), \bibinfo{pages}{1--43}.
\newblock


\bibitem[Wang et~al\mbox{.}(2024)]%
        {wang2024intersectional}
\bibfield{author}{\bibinfo{person}{Yifan Wang}, \bibinfo{person}{Peijie Sun}, \bibinfo{person}{Weizhi Ma}, \bibinfo{person}{Min Zhang}, \bibinfo{person}{Yuan Zhang}, \bibinfo{person}{Peng Jiang}, {and} \bibinfo{person}{Shaoping Ma}.} \bibinfo{year}{2024}\natexlab{}.
\newblock \showarticletitle{Intersectional two-sided fairness in recommendation}. In \bibinfo{booktitle}{\emph{Proceedings of the ACM Web Conference 2024}}. \bibinfo{pages}{3609--3620}.
\newblock


\bibitem[Wei et~al\mbox{.}(2024)]%
        {wei2024llmrec}
\bibfield{author}{\bibinfo{person}{Wei Wei}, \bibinfo{person}{Xubin Ren}, \bibinfo{person}{Jiabin Tang}, \bibinfo{person}{Qinyong Wang}, \bibinfo{person}{Lixin Su}, \bibinfo{person}{Suqi Cheng}, \bibinfo{person}{Junfeng Wang}, \bibinfo{person}{Dawei Yin}, {and} \bibinfo{person}{Chao Huang}.} \bibinfo{year}{2024}\natexlab{}.
\newblock \showarticletitle{Llmrec: Large language models with graph augmentation for recommendation}. In \bibinfo{booktitle}{\emph{Proceedings of the 17th ACM International Conference on Web Search and Data Mining}}. \bibinfo{pages}{806--815}.
\newblock


\bibitem[Wen et~al\mbox{.}(2022)]%
        {wen2022distributionally}
\bibfield{author}{\bibinfo{person}{Hongyi Wen}, \bibinfo{person}{Xinyang Yi}, \bibinfo{person}{Tiansheng Yao}, \bibinfo{person}{Jiaxi Tang}, \bibinfo{person}{Lichan Hong}, {and} \bibinfo{person}{Ed~H Chi}.} \bibinfo{year}{2022}\natexlab{}.
\newblock \showarticletitle{Distributionally-robust recommendations for improving worst-case user experience}. In \bibinfo{booktitle}{\emph{Proceedings of the ACM Web Conference 2022}}. \bibinfo{pages}{3606--3610}.
\newblock


\bibitem[Wu et~al\mbox{.}(2022)]%
        {wu2022multi}
\bibfield{author}{\bibinfo{person}{Haolun Wu}, \bibinfo{person}{Chen Ma}, \bibinfo{person}{Bhaskar Mitra}, \bibinfo{person}{Fernando Diaz}, {and} \bibinfo{person}{Xue Liu}.} \bibinfo{year}{2022}\natexlab{}.
\newblock \showarticletitle{A multi-objective optimization framework for multi-stakeholder fairness-aware recommendation}.
\newblock \bibinfo{journal}{\emph{ACM Transactions on Information Systems}} \bibinfo{volume}{41}, \bibinfo{number}{2} (\bibinfo{year}{2022}), \bibinfo{pages}{1--29}.
\newblock


\bibitem[Wu et~al\mbox{.}(2021)]%
        {wu2021self}
\bibfield{author}{\bibinfo{person}{Jiancan Wu}, \bibinfo{person}{Xiang Wang}, \bibinfo{person}{Fuli Feng}, \bibinfo{person}{Xiangnan He}, \bibinfo{person}{Liang Chen}, \bibinfo{person}{Jianxun Lian}, {and} \bibinfo{person}{Xing Xie}.} \bibinfo{year}{2021}\natexlab{}.
\newblock \showarticletitle{Self-supervised graph learning for recommendation}. In \bibinfo{booktitle}{\emph{Proceedings of the 44th international ACM SIGIR conference on research and development in information retrieval}}. \bibinfo{pages}{726--735}.
\newblock


\bibitem[Xi et~al\mbox{.}(2024)]%
        {xi2024towards}
\bibfield{author}{\bibinfo{person}{Yunjia Xi}, \bibinfo{person}{Weiwen Liu}, \bibinfo{person}{Jianghao Lin}, \bibinfo{person}{Xiaoling Cai}, \bibinfo{person}{Hong Zhu}, \bibinfo{person}{Jieming Zhu}, \bibinfo{person}{Bo Chen}, \bibinfo{person}{Ruiming Tang}, \bibinfo{person}{Weinan Zhang}, {and} \bibinfo{person}{Yong Yu}.} \bibinfo{year}{2024}\natexlab{}.
\newblock \showarticletitle{Towards open-world recommendation with knowledge augmentation from large language models}. In \bibinfo{booktitle}{\emph{Proceedings of the 18th ACM Conference on Recommender Systems}}. \bibinfo{pages}{12--22}.
\newblock


\bibitem[Yang et~al\mbox{.}(2024)]%
        {yang2024qwen2}
\bibfield{author}{\bibinfo{person}{An Yang}, \bibinfo{person}{Baosong Yang}, \bibinfo{person}{Beichen Zhang}, \bibinfo{person}{Binyuan Hui}, \bibinfo{person}{Bo Zheng}, \bibinfo{person}{Bowen Yu}, \bibinfo{person}{Chengyuan Li}, \bibinfo{person}{Dayiheng Liu}, \bibinfo{person}{Fei Huang}, \bibinfo{person}{Haoran Wei}, {et~al\mbox{.}}} \bibinfo{year}{2024}\natexlab{}.
\newblock \showarticletitle{Qwen2. 5 technical report}.
\newblock \bibinfo{journal}{\emph{arXiv preprint arXiv:2412.15115}} (\bibinfo{year}{2024}).
\newblock


\bibitem[Yang et~al\mbox{.}(2023)]%
        {yang2023towards}
\bibfield{author}{\bibinfo{person}{Hao Yang}, \bibinfo{person}{Zhining Liu}, \bibinfo{person}{Zeyu Zhang}, \bibinfo{person}{Chenyi Zhuang}, {and} \bibinfo{person}{Xu Chen}.} \bibinfo{year}{2023}\natexlab{}.
\newblock \showarticletitle{Towards robust fairness-aware recommendation}. In \bibinfo{booktitle}{\emph{Proceedings of the 17th ACM Conference on Recommender Systems}}. \bibinfo{pages}{211--222}.
\newblock


\bibitem[Yeh et~al\mbox{.}(2023)]%
        {yeh2023evaluating}
\bibfield{author}{\bibinfo{person}{Kai-Ching Yeh}, \bibinfo{person}{Jou-An Chi}, \bibinfo{person}{Da-Chen Lian}, {and} \bibinfo{person}{Shu-Kai Hsieh}.} \bibinfo{year}{2023}\natexlab{}.
\newblock \showarticletitle{Evaluating interfaced llm bias}. In \bibinfo{booktitle}{\emph{Proceedings of the 35th Conference on Computational Linguistics and Speech Processing (ROCLING 2023)}}. \bibinfo{pages}{292--299}.
\newblock


\bibitem[Yu et~al\mbox{.}(2023)]%
        {yu2023xsimgcl}
\bibfield{author}{\bibinfo{person}{Junliang Yu}, \bibinfo{person}{Xin Xia}, \bibinfo{person}{Tong Chen}, \bibinfo{person}{Lizhen Cui}, \bibinfo{person}{Nguyen Quoc~Viet Hung}, {and} \bibinfo{person}{Hongzhi Yin}.} \bibinfo{year}{2023}\natexlab{}.
\newblock \showarticletitle{XSimGCL: Towards extremely simple graph contrastive learning for recommendation}.
\newblock \bibinfo{journal}{\emph{IEEE Transactions on Knowledge and Data Engineering}} \bibinfo{volume}{36}, \bibinfo{number}{2} (\bibinfo{year}{2023}), \bibinfo{pages}{913--926}.
\newblock


\bibitem[Zehlike et~al\mbox{.}(2022)]%
        {zehlike2022fairness}
\bibfield{author}{\bibinfo{person}{Meike Zehlike}, \bibinfo{person}{Ke Yang}, {and} \bibinfo{person}{Julia Stoyanovich}.} \bibinfo{year}{2022}\natexlab{}.
\newblock \showarticletitle{Fairness in ranking, part ii: Learning-to-rank and recommender systems}.
\newblock \bibinfo{journal}{\emph{Comput. Surveys}} \bibinfo{volume}{55}, \bibinfo{number}{6} (\bibinfo{year}{2022}), \bibinfo{pages}{1--41}.
\newblock


\bibitem[Zhang et~al\mbox{.}(2024a)]%
        {zhang2024generative}
\bibfield{author}{\bibinfo{person}{An Zhang}, \bibinfo{person}{Yuxin Chen}, \bibinfo{person}{Leheng Sheng}, \bibinfo{person}{Xiang Wang}, {and} \bibinfo{person}{Tat-Seng Chua}.} \bibinfo{year}{2024}\natexlab{a}.
\newblock \showarticletitle{On generative agents in recommendation}. In \bibinfo{booktitle}{\emph{Proceedings of the 47th international ACM SIGIR conference on research and development in Information Retrieval}}. \bibinfo{pages}{1807--1817}.
\newblock


\bibitem[Zhang et~al\mbox{.}(2023a)]%
        {zhang2023chatgpt}
\bibfield{author}{\bibinfo{person}{Jizhi Zhang}, \bibinfo{person}{Keqin Bao}, \bibinfo{person}{Yang Zhang}, \bibinfo{person}{Wenjie Wang}, \bibinfo{person}{Fuli Feng}, {and} \bibinfo{person}{Xiangnan He}.} \bibinfo{year}{2023}\natexlab{a}.
\newblock \showarticletitle{Is chatgpt fair for recommendation? evaluating fairness in large language model recommendation}. In \bibinfo{booktitle}{\emph{Proceedings of the 17th ACM Conference on Recommender Systems}}. \bibinfo{pages}{993--999}.
\newblock


\bibitem[Zhang et~al\mbox{.}(2023b)]%
        {zhang2023recommendation}
\bibfield{author}{\bibinfo{person}{Junjie Zhang}, \bibinfo{person}{Ruobing Xie}, \bibinfo{person}{Yupeng Hou}, \bibinfo{person}{Xin Zhao}, \bibinfo{person}{Leyu Lin}, {and} \bibinfo{person}{Ji-Rong Wen}.} \bibinfo{year}{2023}\natexlab{b}.
\newblock \showarticletitle{Recommendation as instruction following: A large language model empowered recommendation approach}.
\newblock \bibinfo{journal}{\emph{ACM Transactions on Information Systems}} (\bibinfo{year}{2023}).
\newblock


\bibitem[Zhang and Liu(2023)]%
        {zhang2023customized}
\bibfield{author}{\bibinfo{person}{Kaidong Zhang} {and} \bibinfo{person}{Dong Liu}.} \bibinfo{year}{2023}\natexlab{}.
\newblock \showarticletitle{Customized segment anything model for medical image segmentation}.
\newblock \bibinfo{journal}{\emph{arXiv preprint arXiv:2304.13785}} (\bibinfo{year}{2023}).
\newblock


\bibitem[Zhang et~al\mbox{.}(2024b)]%
        {zhang2024blo}
\bibfield{author}{\bibinfo{person}{Li Zhang}, \bibinfo{person}{Youwei Liang}, \bibinfo{person}{Ruiyi Zhang}, \bibinfo{person}{Amirhosein Javadi}, {and} \bibinfo{person}{Pengtao Xie}.} \bibinfo{year}{2024}\natexlab{b}.
\newblock \showarticletitle{BLO-SAM: bi-level optimization based finetuning of the segment anything model for overfitting-preventing semantic segmentation}. In \bibinfo{booktitle}{\emph{Forty-first International Conference on Machine Learning}}.
\newblock


\bibitem[Zhao et~al\mbox{.}(2023)]%
        {zhao2023popularity}
\bibfield{author}{\bibinfo{person}{Jujia Zhao}, \bibinfo{person}{Wenjie Wang}, \bibinfo{person}{Xinyu Lin}, \bibinfo{person}{Leigang Qu}, \bibinfo{person}{Jizhi Zhang}, {and} \bibinfo{person}{Tat-Seng Chua}.} \bibinfo{year}{2023}\natexlab{}.
\newblock \showarticletitle{Popularity-aware distributionally robust optimization for recommendation system}. In \bibinfo{booktitle}{\emph{Proceedings of the 32nd ACM International Conference on Information and Knowledge Management}}. \bibinfo{pages}{4967--4973}.
\newblock


\end{thebibliography}

\clearpage
\appendix

\section{Derivation of Eq.(\ref{equ:w})}
\label{app:derivation}
We provide the derivation for the objective in Eq.(\ref{equ:w}). The goal is to compute the gradient of $H_s(\mathcal{L}(\boldsymbol{\theta}^*(w)))$ with respect to $\boldsymbol{\theta}$, as this will define the optimal direction for updating $w$. We denote $ p_n = \frac{e^{\ell_n(\boldsymbol{\theta}_t)}}{\sum_{i=1}^{N} e^{\ell_i(\boldsymbol{\theta}_t)}} $ and \( d_t(w) \) represents the gradient-weighted update vector under the current weight assignment \( w \). The gradient is computed as:
\begin{align}
    & \nabla H_s(\mathcal{L}(\boldsymbol{\theta}^*(w))) \nonumber\\
    & = - \sum_{n=1}^{N} \left\{ \nabla p_i [\log(p_i) + 1] \right\} \nonumber\\
    & = - \sum_{n=1}^{N} \left\{ [\log(p_i) + 1] \nabla \frac{e^{\ell_i(\boldsymbol{\theta})}}{\sum_{j=1}^{N} e^{\ell_j(\boldsymbol{\theta})}} \right\} \nonumber\\
    & = - \sum_{n=1}^{N} \left\{ [\log(p_i) + 1]
    \frac{\sum_{j=1}^{N} e^{\ell_j(\boldsymbol{\theta})} e^{\ell_i(\boldsymbol{\theta})} \nabla \ell_i(\boldsymbol{\theta})}{(\sum_{j=1}^{N} e^{\ell_j(\boldsymbol{\theta})})^2} \right. \nonumber\\
    & \qquad \left. - [\log(p_i) + 1]
    \frac{e^{\ell_i(\boldsymbol{\theta})} \sum_{j=1}^{N} e^{\ell_j(\boldsymbol{\theta})} \nabla \ell_j(\boldsymbol{\theta})}{(\sum_{j=1}^{N} e^{\ell_j(\boldsymbol{\theta})})^2} \right\} \nonumber\\
    & = - \frac{1}{(\sum_{j=1}^{N} e^{\ell_j(\boldsymbol{\theta})})^2} \sum_{n=1}^{N} \left\{ [\log(p_i) + 1] e^{\ell_i(\boldsymbol{\theta})} \sum_{j=1}^{N} e^{\ell_j(\boldsymbol{\theta})} (\nabla \ell_i(\boldsymbol{\theta}) 
    \right. \nonumber\\
    & \qquad \qquad \left. - \nabla \ell_j(\boldsymbol{\theta})) \right\} \nonumber\\
    & = - \frac{1}{(\sum_{j=1}^{N} e^{\ell_j(\boldsymbol{\theta})})^2} \sum_{n=1}^{N} \left[ \log(p_i) e^{\ell_i(\boldsymbol{\theta})} \sum_{j=1}^{N} e^{\ell_j(\boldsymbol{\theta})} (\nabla \ell_i(\boldsymbol{\theta}) - \nabla \ell_j(\boldsymbol{\theta})) \right] \nonumber\\
    & = - \frac{1}{(\sum_{j=1}^{N} e^{\ell_j(\boldsymbol{\theta})})^2} \sum_{n=1}^{N} \nabla \ell_i(\boldsymbol{\theta}) \left[ \log(p_i) e^{\ell_i(\boldsymbol{\theta})} \sum_{j=1}^{N} e^{\ell_j(\boldsymbol{\theta})} 
    \right. \nonumber\\
    & \qquad \left.- e^{\ell_i(\boldsymbol{\theta})} \sum_{j=1}^{N} \log(p_j) e^{\ell_j(\boldsymbol{\theta})} \right] \nonumber\\
    & = \sum_{n=1}^{N} \nabla \ell_i(\boldsymbol{\theta}) [p_i \sum_{j=1}^{N} \log(p_j) p_j - p_i \log(p_i)]
\end{align}
\section{Experiment details and more results}
\label{app:exp}
\subsection{Datasets}
\label{app:dataset}
Table ~\ref{tab:dataset} provides statistical details of datasets.
\subsection{Brief of Used RSs}
We briefly introduce the RSs we use.

\begin{itemize}[leftmargin=*]\setlength{\itemsep}{2pt}
    \item \textbf{MF}~\cite{koren2009matrix}: is one of the foundational collaborative filtering models, where user and item interactions are represented as ID-based embeddings. It applies matrix factorization with a recommendation loss function.
    \item \textbf{MultVAE}~\cite{liang2018variational}: is based on variational autoencoders (VAE). It frames item recommendation as a generative task modeled by a multinomial distribution and utilizes variational inference for parameter estimation.
    \item \textbf{LightGCN}~\cite{he2020lightgcn}: is a simplified graph convolutional network designed specifically for recommendation systems. It removes unnecessary feature transformations and activation functions present in previous models, enhancing efficiency and performance.
    \item \textbf{SGL}~\cite{wu2021self}: introduces graph contrastive learning to recommendation models. By applying node or edge dropout to create augmented graph views, it performs contrastive learning between two views, surpassing the performance of LightGCN.
    \item \textbf{XSimGCL}~\cite{yu2023xsimgcl}: is a more recent model, generating augmented views by injecting noise into the internal layers of LightGCN, rather than using graph augmentation. Its streamlined approach results in faster convergence and improved overall performance.
    \item \textbf{Linear}~\cite{sheng2024language}: projects language representations into a behavior space for recommendation, training a matrix to bridge the gap between language and recommendation spaces. By using frozen LMs to generate item and user representations from item metadata, the method maps these representations to behavior space with a linear matrix. This approach suggests that collaborative signals might be encoded in the language space.
    \item \textbf{AlphaRec}~\cite{sheng2024language}: is a recommendation model based solely on language representations, without ID-based embeddings. It incorporates components like MLP, graph convolution, and contrastive learning. By using language representations from item metadata, AlphaRec outperforms traditional CF models, showcasing its effectiveness in capturing user intent and providing better zero-shot recommendations.
\end{itemize}
\begin{table}
\centering
\caption{Summary of datasets.}
\label{tab:dataset}
\begin{tabular}{lcccc}  
\toprule
Dataset   & \# Users    & \# Items & \# Interactions & Density\\
\midrule
Movies  & 47523 & 31154 & 441460 & 0.0003
 \\
Games  &  22964 & 18179 & 172006 & 0.0004
 \\
Books  &  28931 & 38936 & 256128 & 0.0002
 \\
\bottomrule
\end{tabular}
\end{table}
\subsection{Brief of Used Base LLMs}
We briefly introduce the Base LLMs we use for LLM-enhanced RSs.
\begin{itemize}[leftmargin=*]\setlength{\itemsep}{2pt}
    \item \textbf{Llama2-7B}~\cite{touvron2023llama}: is an open-source LLM with 7 billion parameters. It uses grouped-query attention, offering a longer context length and a larger pre-training corpus compared to previous version, resulting in improved performance.
    \item \textbf{Qwen-2.5-7B}~\cite{yang2024qwen2}: is another LLM with 7 billion parameters. It enhances the quality of text generation through advanced attention mechanisms and efficient training strategies, providing strong performance across various NLP tasks.
    \item \textbf{SFR-Embedding-Mistral-7B}~\cite{meng2024sfrembedding}: is a LLM-based text embedding model. It's built on the Mistral-7B. It introduces task-homogeneous batching and optimizes contrastive loss achieving better performance than the original LLM.
\end{itemize}
\begin{table*}[t]
\caption{Performance comparison of recommendation models across accuracy, pop. fairness, and genre fairness metrics on Movies Dataset. Bold values indicate the best results, with improvements calculated relative to the best results of traditional RSs. Arrows ($\uparrow$/$\downarrow$) denote the preferred metric direction.}
\label{tab:app-movie}
\centering
\resizebox{\textwidth}{!}{
\begin{tabular}{l|ccc|cc|cc}
    \toprule
    \multicolumn{1}{c|}{\multirow{2}[4]{*}{Model}} & \multicolumn{3}{c|}{Accuracy} & \multicolumn{2}{c|}{Pop. Fairness} & \multicolumn{2}{c}{Genre Fairness} \\
\cmidrule{2-8}          & Recall$\uparrow$ & NDCG$\uparrow$  & HR$\uparrow$    & CV$\downarrow$ & MIN$\uparrow$   & CV$\downarrow$ & MIN$\uparrow$ \\
    \midrule
    MF    & 0.1038 & 0.2529 & 0.0996 & 0.5344 & 0.0390 & 1.4988 & 0.0000 \\
    MultVAE & 0.1449 & 0.3341 & 0.1368 & 0.5276 & 0.0558 & 1.2432 & 0.0067 \\
    LightGCN & 0.1394 & 0.3219 & 0.1285 & 0.5530 & 0.0549 & 1.3048 & 0.0046 \\
    SGL   & 0.1465 & 0.3359 & 0.1289 & 0.6398 & 0.0400 & 1.4547 & 0.0000 \\
    XSimGCL & 0.1508 & 0.3442 & 0.1355 & 0.7513 & 0.0290 & 1.1627 & 0.0000 \\
    \midrule
    LM(qwen) & 0.1574 & 0.3467 & 0.1487 & 0.4236 & 0.0750 & 1.2211 & 0.0034 \\
    LM(llama) & 0.1756 & 0.3810 & 0.1600 & 0.3475 & 0.0965 & 1.0649 & 0.0095 \\
    LM(SFR) & 0.1969 & 0.4194 & 0.1750 & \textbf{0.2673}(-49.3\%) & \textbf{0.1224}(+119.4\%) & 0.9643 & 0.0106 \\
    AlphaRec(qwen) & 0.1733 & 0.3844 & 0.1631 & 0.4755 & 0.0760 & 1.1461 & 0.0087 \\
    AlphaRec(llama) & 0.1852 & 0.4035 & 0.1700 & 0.4198 & 0.0904 & 1.0602 & 0.0099 \\
    AlphaRec(SFR) & \textbf{0.2034}(+34.8\%) & \textbf{0.4338}(+26.0\%) & \textbf{0.1831}(+33.8\%) & 0.3653 & 0.1089 & \textbf{0.9138}(-21.4\%) & \textbf{0.0313}(+367.2\%) \\
    \bottomrule
    \end{tabular}%
}
\end{table*}

\begin{table*}[t]
\caption{Performance comparison of recommendation models across accuracy, pop. fairness, and genre fairness metrics on Games Dataset. Bold values indicate the best results, with improvements calculated relative to the best results of traditional RSs. Arrows ($\uparrow$/$\downarrow$) denote the preferred metric direction.}
\label{tab:app-game}
\centering
\resizebox{\textwidth}{!}{
\begin{tabular}{l|ccc|cc|cc}
    \toprule
    \multicolumn{1}{c|}{\multirow{2}[4]{*}{Model}} & \multicolumn{3}{c|}{Accuracy} & \multicolumn{2}{c|}{Pop. Fairness} & \multicolumn{2}{c}{Genre Fairness} \\
\cmidrule{2-8}          & Recall$\uparrow$ & NDCG$\uparrow$  & HR$\uparrow$    & CV$\downarrow$ & MIN$\uparrow$   & CV$\downarrow$ & MIN$\uparrow$ \\
    \midrule
    MF    & 0.0612 & 0.1495 & 0.0611 & 0.5812 & 0.0227 & 0.7554 & 0.0087 \\
    MultVAE & 0.1051 & 0.2421 & 0.0882 & 0.5985 & 0.0376 & 0.4915 & 0.0331 \\
    LightGCN & 0.1062 & 0.2435 & 0.0887 & 0.7360 & 0.0292 & 0.5941 & 0.0210 \\
    SGL   & 0.1204 & 0.2722 & 0.0927 & 0.6587 & 0.0323 & 0.5346 & 0.0274 \\
    XSimGCL & 0.1253 & 0.2797 & 0.0917 & 0.7718 & 0.0243 & 0.6087 & 0.0169 \\
    \midrule
    Linear(Qwen) & 0.1146 & 0.2562 & 0.0927 & 0.4365 & 0.0596 & 0.5056 & 0.0372 \\
    Linear(Llama) & 0.1451 & 0.3122 & 0.1068 & 0.3242 & 0.0913 & 0.6424 & 0.0635 \\
    Linear(SFR) & 0.1550 & 0.3308 & 0.1120 & \textbf{0.2626}(-54.8\%) & \textbf{0.1064}(+183.0\%) & \textbf{0.3889}(-20.9\%) & 0.0746 \\
    AlphaRec(Qwen) & 0.1324 & 0.2939 & 0.1021 & 0.5210 & 0.0578 & 0.5454 & 0.0267 \\
    AlphaRec(Llama) & 0.1601 & 0.3406 & 0.1159 & 0.4668 & 0.0782 & 0.4152 & 0.0630 \\
    AlphaRec(SFR) & \textbf{0.1750}(+39.7\%) & \textbf{0.3661}(+30.9\%) & \textbf{0.1250}(+34.8\%) & 0.4388 & 0.0894 & 0.3896 & \textbf{0.0788}(+138.1\%) \\
    \bottomrule
    \end{tabular}%
}
\end{table*}

\subsection{More Implement Details}
\begin{itemize}[leftmargin=*]\setlength{\itemsep}{2pt}
    \item \textbf{Recommendation Models}
    For traditional RSs, our goal is to highlight the fairness gap between them and LLM-enhanced RSs. Therefore, we primarily follow the original implementations or perform simple dataset-specific hyperparameter tuning. For MultiVAE, we adopt the architecture from the original paper, with layer sizes set to 600, 200, and 600. The hyperparameters for SGL and XSimGCL are also aligned with those reported in their respective papers. For LLM-enhanced RSs, item representations are generated based on item titles. For AlphaRec, we adopt varying hidden dimensions for the intermediate MLP depending on the underlying LLM, ensuring a consistent total number of model parameters across different input representation sizes.
    \item \textbf{Compared Methods}
    The settings for the compared methods mainly follow their original implementations, with hyperparameter search conducted within the ranges provided in their papers. To ensure a fair comparison, for MultiFR, we adopt group-based loss as the fairness objective. For Reweight, we use the inverse frequency of each item's group, normalized to serve as the weight. For DRORec, the step size for weight updates is set to 0.01. For PDRO, we set the number of stages to 4, the strength of the popularity factor to 3, and the popularity trend strength to 0.3. For ITFR, the perturbation radius for sharpness-aware optimization is set to 0.05.
    \item \textbf{Hyper-parameters}
    We set the batch size to 4096 and the temperature parameter $\tau$ for the InfoNCE loss to 0.1.
\end{itemize}
\subsection{More Results}
We present additional empirical results on the Movies and Games datasets in Table ~\ref{tab:app-movie} and Table ~\ref{tab:app-game}.

\end{document}